\documentclass[12pt]{article}
\usepackage{amsmath}
\usepackage{latexsym}
\usepackage{amssymb}
\usepackage{a4wide}
\usepackage{bbm}
\usepackage{epsfig}
\usepackage{color}
\newcommand{\I}{\text{i}}
\newcommand{\E}{e}

\newcommand{\STr}{\text{STr}}

\newcommand{\be}{\begin{equation}}
\newcommand{\ee}{\end{equation}}
\newcommand{\bear}{\begin{eqnarray}}
\newcommand{\ear}{\end{eqnarray}}

\newcommand{\re}[1]{~(\ref{#1})}
\newcommand{\case}[2]{{\scriptstyle \frac{#1}{#2}}}
\newcommand{\Gk}{\Gamma_k}
\newcommand{\yl}{\psi_{\text{L}}}
\newcommand{\yr}{\psi_{\text{R}}}
\newcommand{\ybl}{\bar{\psi}_{\text{L}}}
\newcommand{\ybr}{\bar{\psi}_{\text{R}}}
\newcommand{\yb}{\bar{\psi}}

\newcommand{\lk}{\bar{\lambda}_{\sigma}}
\newcommand{\mkq}{\bar{m}^2}
\newcommand{\hk}{\bar{h}}
\newcommand{\Zphi}{Z_{\phi}}
\newcommand{\Zy}{Z_{\psi}}
\newcommand{\ZF}{Z_{\text{F}}}
\newcommand{\dlk}{\Delta\lk}

\newcommand{\fss}[1]{#1\!\!\!/}
\newcommand{\fsl}[1]{#1\!\!\!\!/}

\newcommand{\phid}{\phi^\ast}

\newcommand{\pat}{\partial_t}
\newcommand{\patt}{\tilde{\partial}_t}

\newcommand{\te}{\tilde{\epsilon}}

\newcommand{\Nc}{N_{\text{c}}}
\newcommand{\Nf}{N_{\text{f}}}

\newcommand{\SPs}{\,(\text{S--P})_{\text{S}}}
\newcommand{\SPo}{\,(\text{S--P})_{\Nc^2-1}}
\newcommand{\Vs}{\,(\text{V})_{\text{S}}}
\newcommand{\Vo}{\,(\text{V})_{\Nc^2-1}}

\newcommand{\Cas}{C_2(\Nc)}
\newcommand{\xsb}{$\chi$SB\ } 
\newcommand{\gb}{\bar{g}}
\newcommand{\lf}{\bar{\lambda}_{\phi}}
\newcommand{\nub}{\bar{\nu}}
\newcommand{\etaF}{\eta_{\text{F}}}
\newcommand{\rF}{r_{\text{F}}}
\newcommand{\wF}{w_{\text{F}}}
\newcommand{\PF}{P_{\text{F}}}
\newcommand{\wB}{w_{\text{B}}}
\newcommand{\UA}{\text{U}_{\text{A}}(1)}
\newcommand{\rhov}{\varrho}
\newcommand{\alphas}{\alpha_{\text{s}}}
\newcommand{\mIA}{\bar{m}_{\text{I+A}}}
\begin{document}
 
\vspace{-2cm}

{\hfill CERN-TH/2002-242}

{\hfill HD-THEP-02-33}

\vspace{1.5cm}

\centerline{\Large\bf  Universality of spontaneous chiral symmetry
  breaking}

\vspace{2mm}

\centerline{\Large\bf  in gauge theories} 

\vspace{.8cm}

\centerline{\large Holger Gies${}^a$ and Christof Wetterich${}^b$}
 
\vspace{.6cm}

\centerline{\small\it ${}^a$ CERN, Theory Division, CH-1211 Geneva 23,
  Switzerland }
\centerline{\small\it \quad E-mail: Holger.Gies@cern.ch}

\vspace{.1cm}

\centerline{\small\it ${}^b$ Institut f\"ur theoretische Physik,
  Universit\"at Heidelberg,}
\centerline{\small\it Philosophenweg 16, D-69120 Heidelberg,
  Germany} 
\centerline{\small\it \quad E-mail: C.Wetterich@thphys.uni-heidelberg.de}

\begin{abstract}
  We investigate one-flavor QCD with an additional chiral scalar
  field. For a large domain in the space of coupling constants, this
  model belongs to the same universality class as QCD, and the effects
  of the scalar become unobservable. This is connected to a
  ``bound-state fixed point'' of the renormalization flow for which
  all memory of the microscopic scalar interactions is lost. The
  QCD domain includes a microscopic scalar potential with minima at
  nonzero field. On the other hand, for a scalar mass term $m^2$ below
  a critical value ${m}_{\text{c}}^2$, the universality class is
  characterized by perturbative spontaneous chiral symmetry breaking
  which renders the quarks massive. Our renormalization group analysis
  shows how this universality class is continuously connected with the
  QCD universality class. 
\end{abstract}

\section{Introduction}

Universality of QCD means that predictions are independent of the
details of the microscopic interactions. This is crucial for
predictivity, since the precise form of the fundamental interactions
at very short distance scales is not known. In a large parameter space
characterizing possible fundamental interactions, the QCD universality
class corresponds, however, only to a certain domain. For other
domains in parameter space, the color symmetry may be ``spontaneously
broken'' by the Higgs mechanism, or all quarks may acquire a large mass
due to spontaneous chiral symmetry breaking. We are interested here in
the transition from one domain to another and in the question of what
happens at the boundary of the ``QCD domain''.

Looking at QCD from a microscopic scale -- say a unification scale
$10^{15}$GeV -- its universality class is characterized by
eight massless gluons and a certain number of massless fermions.
Perturbatively, the masses are protected by the gauge symmetry and
chiral symmetries. At a much smaller scale around 1GeV, nonperturbative
effects induce masses for all physical particles. In particular, the
fermions become massive owing to chiral symmetry breaking
(\xsb$\!\!$).  This may be described by a nonzero expectation value
$\sigma\sim \langle\yb\psi\rangle$ of a ``composite'' scalar field. In
order to keep the discussion simple, we concentrate here on the case
of one quark flavor -- generalizations to several flavors are
straightforward.

Let us now consider a class of microscopic theories with a complex
fundamental ``chiral scalar field'' $\phi$ which has the same
transformation properties as $\yb\psi$ and a classical potential
\begin{equation}
V=m^2\phi^\ast\phi +\frac{1}{2} \lambda_\phi (\phi^\ast \phi)^2.
\label{A}
\end{equation}
The symmetries also allow for a Yukawa coupling between $\phi$ and the
quarks. For nonzero $\langle \phi\rangle$, the chiral symmetry is
broken and the quarks become massive. In the case of large enough
positive $m^2$ (in units of some unification scale, say $10^{15}$GeV),
the scalar field is super-heavy and decouples from the low-energy
theory. This range of $m^2$ obviously corresponds to the universality
class of QCD. All effects of the scalar field are suppressed by
$p^2/m^2$, with $p$ a characteristic momentum. For QCD predictions,
they can be completely ignored.

On the other hand, for large enough negative $m^2$, we expect the
perturbative picture of spontaneous symmetry breaking to hold. The
scalar field gets a vacuum expectation value (VEV)
\begin{equation}
\langle \phi
\rangle=\sigma=|m_{\text{R}}^2/\lambda_{\phi,\text{R}}|^{1/2},
\label{B}
\end{equation}
with $m_{\text{R}}$ and $\lambda_{\phi,\text{R}}$ related to $m$ and
$\lambda_\phi$ by renormalization corrections. Both $\sigma$ and the
quark masses are of the order of the unification scale in this domain.
The universality class now corresponds to gluodynamics without light
quarks. In the chiral limit of a vanishing current quark mass,
spontaneous \xsb also generates a very light
pseudo-Goldstone boson in addition to the gluonic degrees of freedom.

Varying the microscopic scalar mass term $m^2$ from large negative to
large positive values should lead us from the universality class with
perturbative spontaneous chiral symmetry breaking (P\xsb) to the
universality class of one-flavor QCD. One of the aims of this note is
to understand the qualitative features of this transition in the
vicinity of a critical value $m_{\text{c}}^2$. This is clearly a
nonperturbative problem, since on the QCD side of the transition the
effective gauge coupling grows large. 

Our investigation is based on a nonperturbative flow equation which is
obtained by a truncation of the exact renormalization group equation
for the effective average action \cite{Wetterich:1993yh}. A crucial
ingredient is the ``bosonization'' of effective multi-fermion
interactions at every scale \cite{Gies:2001nw}. This provides for a
description of fundamental scalar fields and bound states in a unified
framework. A theoretical method with this feature is actually required
for our problem, since the scalar quark-antiquark bound states in the
QCD description (e.g., the pseudo-Goldstone eta meson and the
sigma meson) are expected to become associated with the fundamental
scalar in the P\xsb description. In this framework, we see also how
one relevant and two marginal parameters in the P\xsb universality
class, namely the ones corresponding to the mass and quartic
self-interaction of the scalar field and the Yukawa coupling, become
irrelevant for the QCD universality class.

This remarkable change of the number of relevant parameters at the
transition between the two universality classes is connected with the
appearance of a bound-state fixed point for the flow of the scalar
mass and self-interaction in the range of microscopic parameters
corresponding to QCD. This bound-state fixed point is infrared
attractive for all couplings except for the gauge coupling. Under the
influence of this fixed point, all memory of the details of the
microscopic interactions in the scalar sector is lost. This is exactly
what is required for the QCD universality class which has the gauge
coupling as the only marginal parameter (for a massless quark). In
order to see the appearance of the bound state, it is crucial to
re-incorporate the effective multi-fermion interactions generated by
the flow into the effective bosonic interactions. This avoids an
unwanted redundancy of the description. It also solves an old problem
in the investigation of gauged Nambu--Jona-Lasinio models \cite{NJL};
namely, how the presence of apparent relevant parameters in a too naive
treatment of these models can be reconciled with QCD, where no such
relevant parameters are present. In our approach, the flow towards the
bound-state fixed point solves this generic problem.

As a result of our investigation, we find a qualitatively convincing
picture of the transition between the two universality classes
investigated. We have kept the truncation simple in order to
illustrate the change in the number of relevant and marginal
parameters in a simple way. The price to be paid is a limited accuracy
in the quantitative description for parameter regions where the
effective gauge coupling grows large. In our setting, this concerns
primarily the quantitative details of the flow of the
instanton-mediated interactions and the running of the strong gauge
coupling. We emphasize that the qualitative picture does not require a
detailed understanding of strong interactions in the momentum range
where the gauge coupling is large. All decisive features are
determined by the flow in a momentum range substantially larger than
1GeV. In the same spirit, we also have neglected other effective
bosonic degrees of freedom which may correspond to additional bound
states. We keep only the composite scalars and the gluons. When
we proceed with our analysis to the strongly coupled gauge sector, we
do not attempt to compute the gluodynamics, but simply model the strong
interactions with an increasing gauge coupling; for the latter, we use
various examples discussed in appendix \ref{beta}. We do not claim
that our truncation of the gauge sector is sufficient in order to
establish chiral symmetry breaking in QCD. A much more elaborate
analysis would be needed for this purpose. We rather take the
spontaneous symmetry breaking in the QCD universality class as a fact
(established by other methods and observation). We only require that a
reasonable truncation should describe chiral symmetry breaking. Beyond
this, the details of the truncation in the gauge sector are not
relevant for our discussion of universality classes. Despite
these shortcomings, we expect that our quantitative results describe
the right order of magnitude of one-flavor QCD. An impression of the
size of uncertainties can be gained from Table.~\ref{thetable} in
appendix \ref{beta}.
 
\begin{figure}[t]
\begin{picture}(160,50)
\put(-5,0){
\epsfig{figure=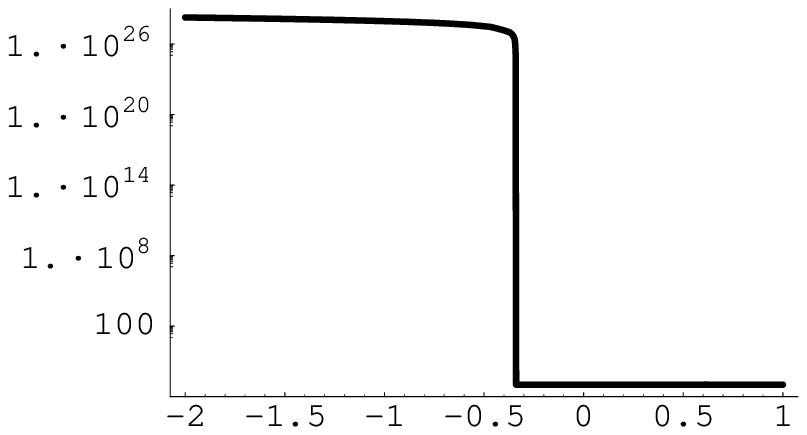,width=7.8cm,height=5cm}}
\put(0,48){(GeV)${}^2$}
\put(50,40){$\sigma_{\text{R}}^2$}
\put(67,0){$\bar{m}_\Lambda^2/\Lambda^2$}
\put(43,0){$\bar{m}_{\text{c}}^2$}
\put(85,3){
\epsfig{figure=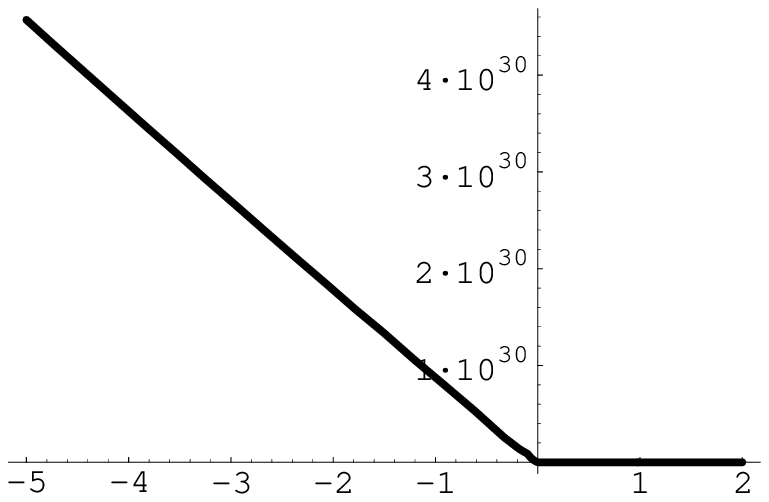,width=7.5cm,height=4cm}}
\put(126,45){(GeV)${}^2$} 
\put(95,28){$\sigma_{\text{R}}^2$}
\put(150,0){$\bar{m}_\Lambda^2/\Lambda^2$}
\put(135,0){$\bar{m}_{\text{c}}^2$}
\end{picture} 


\caption{Renormalized scalar vacuum expectation value
  $\sigma_{\text{R}}^2$ versus the initial condition for the scalar
  mass at the UV cutoff, $\bar{m}_\Lambda^2$. Left panel: logarithmic
  plot for the parameters $Z_\phi=1$, $\bar{h}^2=1$,
  $\bar{\lambda}_\phi=100$ at $\Lambda=10^{15}$GeV resulting in
  $\bar{m}_{\text{c}}^2\simeq-0.35 \Lambda^2$. Right panel: linear
  plot for $\bar{h}^2=0.1$, $\bar{\lambda}_\phi=1$; for
  $\bar{m}^2_{\Lambda}<\bar{m}_{\text{c}}^2\simeq-0.0043\Lambda^2$,
  the linear dependence
  $\sigma_{\text{R}}^2\sim-(\bar{m}^2_{\Lambda}-\bar{m}_{\text{c}}^2)$ 
  as expected from perturbation theory is confirmed.}
\label{fighit} 
\end{figure}

In order to illustrate our points, we compute the scalar condensate,
i.e., the renormalized minimum of the effective potential,
$\sigma_{\text{R}}=\sqrt{Z_\phi |\phi_0|^2}$, for a broad range of
initial scalar mass values $\bar{m}_{\Lambda}^2$. We note that
$\sigma_{\text{R}}$ is directly connected with the decay constant of
the eta meson and sets the scale for the quark mass generated by
\xsb$\!$. We first neglect the anomalous $\UA$ violating contributions
from instanton effects which only affect the physics at scales around
1GeV. (They will be considered in Sect.~\ref{anomaly}.) We parametrize
the microscopic interactions by the initial values of the
renormalization flow at a GUT-like scale $\Lambda=10^{15}$GeV.  As can
be read off from Fig.~\ref{fighit}, a critical mass
$\bar{m}_{\text{c}}^2$ exists. For initial scalar masses below this
critical mass, $\bar{m}_\Lambda^2<\bar{m}_{\text{c}}^2$, the naive
expectation is fulfilled, and we find scalar condensates of the order
of the cutoff, $\sigma_{\text{R}}\sim 10^{13}\dots 10^{15}$GeV. It is
remarkable that the value of the critical mass is negative and
typically of the order of the cutoff or only a few orders of magnitude
below the cutoff; for example, we find
$\bar{m}_{\text{c}}^2\simeq-0.35 \Lambda^2$ for the initial values
$\bar{h}^2=1$ and $\bar{\lambda}_\phi=100$ at $\Lambda=10^{15}$
(Fig.~\ref{fighit} (left panel)). For a perturbatively accessible set
of initial parameters $\bar{h}^2=0.1$ and $\bar{\lambda}_\phi=1$ at
$\Lambda=10^{15}$, we find $\bar{m}_{\text{c}}^2\simeq-0.0043
\Lambda^2$ (Fig.~\ref{fighit} (right panel)). In the latter case, we
find a linear dependence of the condensate on the mass parameter,
$\sigma_{\text{R}}^2\sim -(\bar{m}_\Lambda^2-\bar{m}_{\text{c}}^2)$,
as expected from perturbation theory (cf.~Eq.\re{B}).

However, for initial scalar masses above this critical mass,
$\bar{m}_\Lambda^2>\bar{m}_{\text{c}}^2$, the scalar condensate is 16
orders of magnitude smaller (not visible in the linear plot in
Fig.~\ref{fighit} (right panel)). In this case, symmetry breaking is
triggered by the fermion and gauge sectors and not by the scalar
sector, i.e., $\sigma_{\text{R}}$ is roughly of the order of
$\Lambda_{\text{QCD}}$. Therefore, even if we start the flow deep in
the broken regime with $\bar{m}_\Lambda^2<0$ but above the critical
mass, the scalar fluctuations drive the system first into the
symmetric regime where it will be attracted by the same IR fixed point
as a QCD-like system. It should be stressed that no fine-tuning of the
initial parameters is needed, neither to put the system into the
domain of attraction of the QCD universality class nor to separate the
UV scale from the scale of chiral symmetry breaking.
 
Only for $\bar{m}_\Lambda^2<\bar{m}_{\text{c}}^2$ is the effective
coupling between the scalars and the fermions strong enough to induce
P\xsb with a magnitude determined by the initial parameters of the
scalar sector. In this case, we would have to fine-tune the initial
condition for $\bar{m}_\Lambda^2$ to lie extremely closely to
$\bar{m}_{\text{c}}^2$, if we wanted to separate the UV scale from the
scale of chiral symmetry breaking. This is the famous naturalness
problem which is generic for models involving a fundamental scalar. Of
course, theories without fundamental scalars such as QCD do not have
this problem, although effective scalar degrees of freedom such as
bound states can occur at low energies. It is one of our main
observations that the mechanism of how ``QCD-like'' theories
circumvent the naturalness problem can also be applied to models with
a fundamental scalar.

The details of our study are organized as follows: in
Sect.~\ref{setting}, we introduce the class of models containing
one-flavor QCD and derive the flow equations for a qualitatively
reliable truncation including ``bosonization at all scales''. Section
\ref{bsfp} is devoted to a discussion of the bound-state fixed point
which governs the flow of the QCD domain for weak gauge coupling. In
Sect.~\ref{numerics}, we analyze the universal features of the QCD
domain numerically and give estimates of IR observables in the
nonperturbative strong-coupling regime. Instanton-mediated
interactions are included in Sect.~\ref{anomaly} where we also
describe the fate of the pseudo-Goldstone boson.

\section{Flow equations}
\label{setting}

QCD with one massless Dirac fermion flavor coupled to an SU($\Nc$)
gauge field is characterized by the classical (or bare) action
\begin{equation}
S_{\text{QCD}} 
=\int d^4x\, \yb\, \I \fsl{D}\, [A]\, \psi+\frac{1}{4} F_{\mu\nu}^a
F_{\mu\nu}^a, \label{1.1}
\end{equation}
where $D^{ij}_\mu [A]= \partial_\mu\delta^{ij} -\I \gb T_a^{ij}
A_\mu^a$, and $T_a$ denotes the (hermitean) generators of the gauge
group in the fundamental representation. In this work, we embed
one-flavor QCD in a larger class of chirally invariant theories
including a color-singlet scalar field. For this, we consider the
action 
\begin{eqnarray}
\Gamma&=&\int\biggl\{\Zy\yb \I \fsl{D}\, \psi + \frac{\lk}{2}
\bigl[(\yb\psi)^2 -(\yb\gamma_5\psi)^2\bigr]  \nonumber\\
&&\qquad +\Zphi\partial_\mu\phid\partial_\mu\phi+U(\phi) +\hk
\bigl[ (\ybr\yl)\phi-(\ybl\yr) \phid \big] \nonumber\\
&&\qquad +\frac{\ZF}{4}  F_{\mu\nu}^aF_{\mu\nu}^a + \frac{1}{2\xi}
{(\bar{D}_\mu a_\mu^a)^2} \biggr\},
 \label{1.2}
\end{eqnarray}
{which represents a simple truncation of the space of action functionals
and serves as the basis of our approximations.}  Here we have used the
shorthand $(\yb\psi)=\yb^i\psi_i$ for the color indices. We included a
{background gauge fixing term with parameter $\xi$, and
$A_\mu=\bar{A}_\mu+a_\mu$, $\bar{A}_\mu$ being the background and
$a_\mu$ the fluctuation field, $\bar{D}_\mu\equiv
D_\mu[\bar{A}]$. Furthermore, }we do not display the ghost sector for
simplicity. Equation \re{1.2} reduces to one-flavor QCD if we set the
four-fermion and the Yukawa interaction equal to zero, $\lk=\hk=0$,
let the scalar field be auxiliary, $Z_\phi=0$, and set $Z_F=1=\Zy$
(the scalar potential is of no importance then). Furthermore, there is
a redundancy in Eq.\re{1.2}: we can compensate for a shift in $\lk$ by
readjusting the Yukawa coupling and the scalar potential corresponding
to a Hubbard-Stratonovich transformation (partial bosonization). But
apart from this redundancy, which will be removed later on by
``re-bosonization'', different initial values for the various
parameters in Eq.\re{1.2} generally correspond to different quantum
theories. Some of these theories will belong to the same universality
class sharing the same low-energy properties, which makes them
indistinguishable from a low-energy physicist's point of view.

We analyze this class of theories in a Wilsonian spirit upon
integrating out quantum fluctuations momentum shell by momentum
shell. For this we employ the formalism based on the exact
renormalization group flow equation for the effective average action
\cite{Wetterich:1993yh}, \cite{Berges:2000ew}, 
\begin{equation}
\pat\Gk=\frac{1}{2}\, \STr\, \Bigl[\pat R_k \bigl( \Gk^{(2)} +R_k
\bigr)^{-1} \Bigr], \label{floweq}
\end{equation}
where $\Gk^{(2)}$ denotes the second functional derivative of the
effective average action $\Gk$ that governs the dynamics of the system
at a momentum scale $k$. The logarithmic scale parameter $t$ is given
by $t=\ln k/\Lambda$, $\pat=k (d/dk)$, where $\Lambda$ denotes the
ultraviolet (UV) scale at which we define the bare action
$\Gamma_\Lambda$. The cutoff function $R_k$ is to some extent
arbitrary and obeys a few restrictions \cite{Berges:2000ew} which
ensure that the flow is well defined and interpolates between the bare
action in the UV and the full quantum effective action $\Gamma_{k\to
  0}$ in the infrared (IR).

We solve the flow equation\re{floweq} by using Eq.\re{1.2} as a
truncation of the space of all possible action functionals. As a
consequence, we promote all couplings and wave function
renormalizations occurring in Eq.\re{1.2} to $k$-dependent
quantities. Although the truncation \re{1.2} represents only a small
subclass of possible operators generated by quantum fluctuations, it
is able to capture many physical features of QCD-like systems. 

Let us elucidate the single components in detail: for the scalar
potential, we use the simple truncation
\begin{eqnarray}
U(\phi)&=&\mkq\, \rho+ \frac{1}{2} \lf\, \rho^2 -\frac{1}{2} \nub\,
\zeta,\qquad \rho=\phid\phi, \quad \zeta =\phi+\phid. \label{1.3} 
\end{eqnarray}
Already the $\rho$-dependent first two terms of the potential are
capable of describing spontaneous \xsb of the system which we are
aiming at. Indeed, the order parameter $\sigma$ denotes the minimum of
the scale-dependent effective potential $U_k$ for $k\to 0$. The term
$\sim \zeta=\phi+\phid$ breaks the $\text{U}_{\text{A}}(1)$ symmetry
of simultaneous axial phase rotations of scalars and fermions; it
accounts for the effects of the axial anomaly. However, the presence
of the axial anomaly is not relevant for {universality}
of spontaneous \xsb, although it has, of course, a strong quantitative
impact on resulting low-energy parameters such as condensates and
constituent quark masses. Therefore, we postpone the discussion of
this quantitative influence to Sect.~\ref{anomaly} and set $\nub=0$ in
the following for the sake of clarity.

In the gauge sector, we do not attempt to calculate the full
nonperturbative flow of $\ZF$, or alternatively the gauge coupling
$g$, here, but study various possibilities for these flows and take
over nonperturbative results from the literature. The most important
features of the universality classes involve only the perturbative
running of $g$.\footnote{The running of g is universal in two-loop
order. In the framework of the exact renormalization group, this has
been computed in \cite{Pawlowski:2001df}. As discussed in the
introduction, the one-loop running is actually sufficient to generate
the main qualitative features needed for our argument.}

We will define the quantum theories by fixing the initial conditions
for the renormalization flow at the UV scale $\Lambda$. In
the gauge and fermion sectors, we choose
\begin{equation}
\ZF\big|_{k=\Lambda}=1, \quad\Zy\big|_{k=\Lambda}=1,\quad
\bar{\lambda}_{\sigma}\big|_{k=\Lambda}=0.  
\label{1.4}
\end{equation}
The first two conditions normalize the gauge and fermion fields and
imply that $\gb$ denotes the bare gauge coupling. The last condition
states that four-fermion interactions either have been partially
bosonized into the scalar sector or are completely absent at the UV
cutoff scale $\Lambda$. 

The choice of the scalar couplings at the UV cutoff will finally
determine whether we are in or beyond the QCD domain. In order to
describe standard QCD in our picture, a natural choice is given by
\begin{equation}
\mkq\big|_{k=\Lambda}=+{\cal O}(\Lambda^2),
\quad \bar{\lambda}_{\phi}\big|_{k=\Lambda}=0,\quad
(\Zphi,\bar{h})\big|_{k=\Lambda}\to 0, 
\label{1.5}
\end{equation}
implying that the scalar fields are nondynamic, noninteracting and
heavy at $\Lambda$ and decouple from the fermion sector. They could be
integrated out without any effect on the fermion sector and therefore
are completely auxiliary. However, we will demonstrate below that the
infrared physics including \xsb is to a large extent independent of
the initial values in the scalar sector; in other words, the QCD
universality class is actually much bigger than the restrictive choice
of initial conditions of Eq.\re{1.5}.\footnote{Already at this point,
  it is clear that $\lambda_{\phi,\Lambda}$ could also be chosen
  nonzero, which would only result in an unimportant change of the
  normalization of the functional integral.}

For a concise presentation of the RG flow equations of the single
couplings, it is convenient to introduce the dimensionless,
renormalized and $k$-dependent quantities,
\begin{equation}
\epsilon=\frac{\mkq}{Z_{\phi}k^2}, \quad
\lambda_\phi=\frac{\lf}{\Zphi^2},\quad
h=\frac{\hk}{Z_{\phi}^{1/2}\Zy}, \label{1.6}
\end{equation}
in the symmetric regime of the system. In the \xsb regime, the mass term
becomes negative, and we replace this coupling by the minimum of the
potential $\rho_0$ and its corresponding dimensionless variable
$\kappa$ defined by 
\begin{equation}
0=\frac{\partial}{\partial \rho} U_k(\rho=\rho_0), \quad
\kappa=\frac{\Zphi\, \rho_0}{k^2}. \label{1.7}
\end{equation}
Similarly, we define $\lf$ as the second $\rho$-derivative of the
potential at the minimum in the \xsb regime. The running of the wave
function renormalizations is studied using the associated anomalous
dimensions,
\begin{equation}
\eta_\phi=-\pat \ln \Zphi, \quad \eta_\psi=-\pat \ln \Zy, \quad
\etaF=-\pat \ln\ZF, \label{1.8}
\end{equation}
{where $\etaF$ represents the major piece of information from the gauge
sector in our truncation. Here, the use of the background-field method
for this gauge sector has two advantages: first, it represents a
book-keeping device to set up consistent gauge-invariant
approximations within a certain order of truncation. Second, the
physical idea of the background field is that it accommodates the true
ground state of the system around which the quantum fluctuations are
integrated out. In this spirit, we deduce the running gauge coupling
from the RG behavior of the background field. Owing to background
gauge invariance, the product of gauge coupling and background gauge
field is renormalization-group invariant \cite{Abbott:1980hw}, so that
the beta function for the renormalized running gauge coupling $g$ is
related to $\etaF$ by 
\begin{equation} \beta_{g^2}\equiv \pat
g^2\,\,=\,\,\etaF\, g^2, \quad g^2=\frac{\gb^2}{\ZF}. \label{1.9}
\end{equation} 
Actually, the effective action depends on both the background and the
fluctuating gauge field, and the $n$-point functions can only be
extracted from the functional depending on both fields
\cite{Reuter}. Nevertheless, once all fluctuations are integrated out,
the fluctuating field can be set to zero and the resulting effective
action is gauge invariant. In general, the dependence of the effective
action on both fields is needed for the RG flow. With the help of
background-field identities, the dependence of the effective action on
the fluctuating gauge field and the background field are related. A
detailed record of the flow equations and results in the background
field formalism, including the role of the gauge symmetry and
Slavnov-Taylor identities, can be found in \cite{Reuter,Bonini}.

In the present work, we neglect possible differences between the RG
flow for gauge couplings defined from the background-field effective
action and from vertices of the fluctuating field
\cite{Pawlowski:2001df,Litim:2002ce}. This is perfectly justified in
the limit of small gauge coupling which is of primary importance for
this work. Here the lowest-order running is universal. By contrast, in
the region of large coupling, our truncation of the gauge sector would
anyway not be reliable if taken at face value, so that we abstain from
resolving the gauge field running and $\etaF$ in this regime. In this
region, we simply model the running of the gauge coupling in order to
obtain a first glance at the \xsb regime. Thereby, we assume that the
influence of higher gluonic operators can be effectively accounted for
by the increase of the gauge coupling. Although this certainly
represents an oversimplification, let us stress that the details of
the flow in the gauge sector are only of secondary importance for the
issue addressed in this paper.}

Inserting the truncation\re{1.2} into the exact RG flow equation for
the effective average action, we find the following results. The
scalar and fermion anomalous dimensions can be written as
\begin{eqnarray}
\eta_\phi&=&4 v_4\, \kappa\lambda_\phi^2\,
   m_{2,2}^4(0,2\kappa\lambda_\phi;\eta_\phi) \nonumber\\
&&+4\Nc v_4\, h^2\left[ m_4^{(\text{F}),4}(\kappa h^2; \eta_\psi)+
   \kappa h^2\, m_2^{(\text{F}),4}(\kappa h^2; \eta_\psi)\right],
   \label{1.10}\\
\eta_\psi&=&2\Cas v_4\, g^2\Bigl[ (3-\xi)\,
   m_{1,2}^{(\text{FB}),4}(\kappa h^2,0;\eta_\psi,\etaF)- 3(1-\xi)\,
   \tilde{m}_{1,1}^{(\text{FB}),4}(\kappa h^2,0;\eta_\psi,\etaF) \Bigr]
\nonumber\\
&&+v_4\, h^2 \bigl[m_{1,2}^{(\text{FB}),4}(\kappa h^2,\epsilon+2\kappa
   \lambda_\phi;\eta_\psi,\eta_\phi) 
   +m_{1,2}^{(\text{FB}),4}(\kappa h^2,\epsilon;\eta_\psi,\eta_\phi)
   \bigr], \label{1.11}
\end{eqnarray}
where $v_4=1/(32\pi^2)$ and $\Cas=(\Nc^2-1)/(2\Nc)$. This
representation is valid in the symmetric as well as in the \xsb regime.
In the former, $\kappa$ has to be set equal to zero, whereas
$\epsilon=0$ has to be chosen in the latter. The various quantities
denoted by $m$ are threshold functions which control the decoupling of
massive modes for decreasing $k$; they also contain all dependencies
on the precise choice of the cutoff function $R_k$. Their definitions
and explicit representations can be found in App.~\ref{threshold} or
in \cite{Berges:2000ew}. 

Equation\re{1.10} agrees with \cite{Berges:2000ew} and
\cite{Jungnickel:1996fp}. We also find agreement for the second line
of Eq.\re{1.11}, whereas the first line arises from the gauge-field
sector (which has not been dealt with in
\cite{Berges:2000ew},\cite{Jungnickel:1996fp}). As a further check, we
note that in the perturbative small-coupling limit, where the
threshold functions $m$ occurring above universally reduce to 1, we
obtain
\begin{equation}
\eta_\phi\big|_{\text{pert.}}=\frac{\Nc}{8\pi^2}\, h^2, \quad 
\eta_\psi\big|_{\text{pert.}}=\xi\, \frac{\Cas}{8\pi^2} \, g^2+
\frac{1}{16\pi^2} \, h^2, \label{1.12}
\end{equation}
which agrees with the literature \cite{Chivukula:1992pm}. 

In the symmetric regime, the flow of the purely scalar sector can be
summarized by
\begin{eqnarray}
\pat\epsilon&=&-(2-\eta_\phi)\epsilon-8v_4\, \lambda_\phi\,
   l_1^4(\epsilon;\eta_\phi) +8\Nc v_4\, h^2\,
   l_1^{(\text{F}),4}(0;\eta_\psi), \label{1.13}\\
\pat\lambda_\phi&=& 2\eta_\phi\, \lambda_\phi+20v_4\, \lambda_\phi^2\, 
   l_2^4(\epsilon;\eta_\phi) -8\Nc v_4\, h^4\,
   l_2^{(\text{F}),4}(0;\eta_\psi), \label{1.14}
\end{eqnarray}
whereas in the \xsb regime, we find
\begin{eqnarray}
\pat\kappa&=&-(2+\eta_\phi)\kappa +2v_4\, 
   l_1^4(0;\eta_\phi)+6v_4\, l_1^4(2\kappa\lambda_\phi;\eta_\phi) 
   -8\Nc v_4\, \frac{h^2}{\lambda_\phi}\, 
    l_1^{(\text{F}),4}(\kappa h^2;\eta_\psi), \label{1.15}\\
\pat\lambda_\phi&=& 2\eta_\phi\, \lambda_\phi
   +2v_4\, \lambda_\phi^2\, l_2^4(0;\eta_\phi) 
   +18v_4\, \lambda_\phi^2\, l_2^4(2\kappa\lambda_\phi;\eta_\phi) 
   -8\Nc v_4\, h^4\, l_2^{(\text{F}),4}(\kappa h^2;\eta_\psi), 
   \label{1.16}
\end{eqnarray}
in complete agreement with the results of \cite{Jungnickel:1996fp}.
Again, the quantities denoted by $l$ are threshold functions
\cite{Berges:2000ew}, \cite{Hoefling:2002hj}. Now we turn to the flow
of the Yukawa coupling, which is driven by all sectors of the system:
\begin{eqnarray}
\pat h^2&=&(2\eta_\psi+\eta_\phi)\, h^2 -4v_4\, h^4 
   \bigl[ l_{1,1}^{(\text{FB}),4}(\kappa h^2,
   \epsilon;\eta_\psi,\eta_\phi) - l_{1,1}^{(\text{FB}),4}(\kappa h^2,
   \epsilon+2\kappa\lambda_\phi;\eta_\psi,\eta_\phi)\bigr] \nonumber\\
&&-8(3+\xi)\Cas v_4\, g^2 h^2\, l_{1,1}^{(\text{FB}),4}(\kappa h^2,
   0;\eta_\psi,\etaF), \label{1.17}
\end{eqnarray}
where we have to set $\kappa=0$ ($\epsilon=0$) in the symmetric
(\xsb$\!\!$) regime. As a check, we take a look at the perturbative
limit,
\begin{equation}
\pat h^2\big|_{\text{pert.}}=\frac{\Nc+1}{8\pi^2}\, h^4
-\frac{3\Cas}{4\pi^2}\, g^2 h^2, \label{1.17p}
\end{equation}
where we rediscover known results and also observe that the
gauge-parameter $\xi$-dependence has dropped out as it should.

A crucial ingredient is the flow of the fermion self-interaction,
which -- in dimensionful representation -- can be written as
\begin{eqnarray}
\pat\lk&=& \frac{\Zy^2}{k^2}\bigl[ \beta_{\lk}^{g^4}\, g^4 
  + \beta_{\lk}^{h^4}\, h^4\bigr], \label{1.18}\\
&&\beta_{\lk}^{g^4}:=-6\,\frac{(\Nc+2)(\Nc-1)}{\Nc^2}\, \Cas  
  \, v_4\, \tilde{l}_{1,2}^{(\text{FB}), 4}(\kappa
  h^2,0;\eta_\psi,\etaF),    \nonumber\\
&&\beta_{\lk}^{h^4}:=\left(\frac{2}{\Nc}+1\right)\, v_4\, 
  \tilde{l}_{1,1,1}^{(\text{FBB}), 4}(\kappa
  h^2,\epsilon,\epsilon+2\kappa\lambda_\phi;\eta_\psi,\eta_\phi).
  \nonumber
\end{eqnarray}
Here we neglected terms $\sim \kappa$ which arise only in the broken
regime but are suppressed therein owing to simultaneously occurring
threshold functions (these terms are similar to the last term in
square brackets in Eq.\re{1.10}, which has hardly any effect on the
results either). In Eq.\re{1.18} as well as in all equations above, we
neglected terms of order $\lk$ on the RHS, because $\lk=0$ will
finally be guaranteed on all scales as discussed below. Furthermore,
we have chosen the same Fierz transformations in the Dirac algebra as
in \cite{Gies:2001nw} and decomposed the possible color structures of
the four-fermion interaction into a color singlet $\SPs$ and color
$\Nc^2-1$-plets $\SPo,\Vo$. In the present work, we focus on the
$\SPs$ term; in principle, the $\Vo$ term could be absorbed into a
$k$-dependent transformation of the nonabelian gauge field in the same
way as suggested in \cite{Gies:2001nw} for the abelian
case.\footnote{By neglecting some of the four-fermion interactions,
  our quantitative result will depend slightly on the choice of the
  Fierz decomposition. Using ``fermion-boson translation'' to be
  described in the following, this dependence can be removed in a
  larger truncation, as was recently shown in \cite{Jaeckel:2002rm}.
  However, we checked explicitly that quantitative results in another
  natural Fierz decomposition involving $\SPs,\Vs$ and $\Vo$ differ
  from the present ones only on the $1\%$ level.}

As mentioned above, there is a certain redundancy in the
parametrization of the effective action $\Gk$ owing to possible
different choices of partial bosonization of the four-fermion
interaction. {From a different viewpoint, this redundancy
corresponds to the possible mixing of fields or composite operators
with identical quantum numbers.} We remove this redundancy in the
present truncation with the aid of the following $k$-dependent
transformation of the scalar field (``fermion-boson translation''):
\begin{eqnarray}
\pat\phi_k(q)&=&-(\ybl\yr)(q)\, \pat\alpha_k(q)+ \phi_k(q)\,
\pat\beta_k(q), \nonumber\\
\pat\phid_k(q)&=&(\ybr\yl)(-q)\, \pat\alpha_k(q)+ \phid_k(q)\,
\pat\beta_k(q),\label{d13ba}
\end{eqnarray}
with a priori arbitrary functions $\alpha_k(q)$ and $\beta_k(q)$. Upon
this transformation, the flow equations given above receive additional
contributions $\sim \alpha_k(q),\beta_k(q)$ according to 
\begin{equation}
\pat\Gk=\pat \Gk{}_{|\phi_k,\phi^\ast_k}+\int \frac{\delta \Gk}{\delta
  \phi_k}\, \pat \phi_k + \int \frac{\delta \Gk}{\delta \phid_k}\, \pat
  \phid_k. \label{cross}
\end{equation}
As described in more detail in \cite{Gies:2001nw}, these functions can
be uniquely determined by demanding for (i) $\pat \lk(q^2)$ to vanish
for all $k$ and $q^2$, where the momentum dependence of $\lk$ has been
studied in the $s$ channel for simplicity, $\lk(q^2)\equiv\lk(s=q^2)$,
(ii) the Yukawa coupling $\hk$ to be momentum independent, and (iii)
$\pat \Zphi(q^2=k^2)=-\eta_\phi \Zphi$ in order to render the
approximation of a momentum-independent $\Zphi$ self-consistent.
Condition (i) together with the initial condition\re{1.4} guarantees
that no four-fermion interaction of this type is generated under the
flow; this interaction is bosonized into the scalar sector at all
scales $k$. Condition (ii) guarantees the fermion mass generated by
\xsb is also momentum independent, so that the couplings in the \xsb
regime have a direct physical interpretation.

The field transformation\re{d13ba} affects also the scalar couplings,
and we obtain in the symmetric regime:
\begin{eqnarray}
\pat\epsilon&=&\pat\epsilon\big|_{\phi_k}
   +2\frac{\epsilon(1+\epsilon)}{h^2}
   \bigl(1+(1+\epsilon)Q_\sigma\bigr) 
   \bigl(\beta_{\lk}^{g^4} \, g^4+\beta_{\lk}^{h^4}\, h^4\bigr),
   \nonumber\\
\pat h^2&=&\pat h^2\big|_{\phi_k}+2\bigl(1+2\epsilon+Q_\sigma
   (1+\epsilon)^2 \bigr) 
   \bigl(\beta_{\lk}^{g^4} \, g^4+\beta_{\lk}^{h^4}\,
   h^4\bigr), \label{1.19} 
\end{eqnarray}
where the corresponding first terms on the right-hand sides denote
the flow equations for fixed fields as given above in Eqs.\re{1.13}
and\re{1.17}. In the \xsb regime, we find similarly
\begin{eqnarray}
\pat\kappa&=&\pat\kappa\big|_{\phi_k}
   +2\frac{\kappa(1-\kappa\lambda_\phi)}{h^2}
   \bigl(1+(1-\kappa\lambda_\phi)Q_\sigma\bigr) 
   \bigl(\beta_{\lk}^{g^4} \, g^4+\beta_{\lk}^{h^4}\, h^4\bigr),
   \nonumber\\
\pat h^2&=&\pat h^2\big|_{\phi_k}+2\bigl(1-2\kappa\lambda_\phi
   +Q_\sigma (1-\kappa\lambda_\phi)^2 \bigr) 
   \bigl(\beta_{\lk}^{g^4} \, g^4+\beta_{\lk}^{h^4}\,
   h^4\bigr). \label{1.19b} 
\end{eqnarray}
Defining $\dlk:=\lk(k^2)-\lk(0)$, the quantity
$Q_\sigma\equiv \pat\dlk/\pat\lk(0)$ measures the suppression of
$\lk(s)$ for large external momenta. Without an explicit computation,
we may conclude that this suppression implies $Q_\sigma<0$, in
agreement with unitarity; furthermore, if the flow is in the \xsb
regime, the fermions become massive, and non-pointlike four-fermion
interactions in the $s$ channel will be suppressed by the inverse
fermion mass squared.\footnote{This can be inferred from the
  heavy-fermion limit of the two-gluon/scalar-exchange box diagram
  where the internal fermion propagators become pointlike $\sim
  1/m_{\text{f}}$.} Therefore, we model $Q_\sigma$ by the ansatz
\begin{equation}
Q_\sigma=Q_\sigma^0\, m_{1,2}^{(\text{FB}),4}(\kappa
h^2,0,\eta_\psi,\eta_F), \quad Q_\sigma^0=\text{const.}<0,
\end{equation}
where we have introduced a threshold function with the appropriate
decoupling properties for massive fermions. The qualitative results
are independent of the precise choice of $Q_\sigma$, and it is
reassuring to observe a quantitative independence of the IR
observables on the precise value for $Q_\sigma^0$ (e.g.,
$Q_\sigma^0\simeq-0.1$). 

The field transformations\re{d13ba} also modify the equation for
$\lambda_\phi$ via the terms $\sim \pat\beta_k$. In the pointlike
limit ($q^2=0$), the modified running is given by
\begin{equation}
\pat\lambda_\phi=\pat\lambda_\phi\big|_{\phi_k}
+4\frac{\lambda_\phi}{h^2} (1+\epsilon)
\bigl(1+(1+\epsilon)Q_\sigma\bigr) \bigl(\beta_{\lk}^{g^4} \,
g^4+\beta_{\lk}^{h^4}\, h^4\bigr).\label{1.20}
\end{equation}
It will turn out that the modification of the flow of $\lambda_\phi$
is also quantitatively irrelevant, whereas the the modifications
displayed in Eqs.\re{1.19} and\re{1.19b} are of crucial importance.

\section{Bound-state fixed point}
\label{bsfp}

The universal features of spontaneous \xsb in the QCD domain that will
be quantitatively analyzed in the next section can be traced back to
the occurrence of a fixed point for the scalar couplings. This fixed
point is infrared attractive as long as the gauge coupling remains
weak and can be associated with a bound state \cite{Gies:2001nw}.

\begin{figure}[t]
\begin{picture}(160,50)
\put(25,-30){
\epsfig{figure=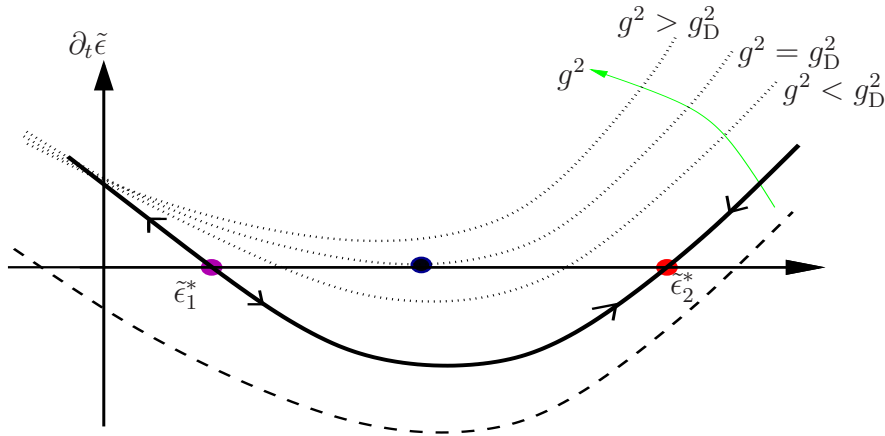,width=11cm,height=12cm}}
\put(49,21){$\te_1^\ast$}
\put(115,22){$\te_2^\ast$}
\put(35,54){$\pat\te$}
\put(100,50){$g^2$}
\put(108,57){$g^2>g_{\text{D}}^2$}
\put(124,53){$g^2=g_{\text{D}}^2$}
\put(130,48){$g^2<g_{\text{D}}^2$}
\end{picture} 


\caption{Flow of $\te$ according to Eq.\re{bs2} (schematic plot): the
  solid line corresponds to a QCD scenario at weak gauge coupling; the
  arrows indicate the direction of the flow towards the infrared. The
  dashed line corresponds to a system with fundamental scalar,
  $Z_\phi|_{k=\Lambda}=1$, $\epsilon\lesssim 1$, and strong scalar
  self-interaction. The dotted lines exhibit the destabilization of
  the bound-state fixed point by the increasing gauge coupling.}
\label{figFP} 
\end{figure}

The fixed-point structure can conveniently be analyzed with the help
of the coupling 
\begin{equation}
\te=\frac{\epsilon}{h^2}=\frac{\Zy^2\mkq}{k^2 \hk^2}. \label{bs1}
\end{equation}
Since we are interested in the domain of weak gauge coupling, for
simplicity we can neglect the anomalous dimensions in the following.
In this approximation and {choosing the gauge parameter $\xi=0$
(background Landau gauge)}, the flow of $\te$ yields:
\begin{eqnarray}
\pat\te&=&8\Nc v_4l_1^{(\text{F}),4} -8 v_4l_1^4(\epsilon)\,
\frac{\lambda_\phi}{h^2} -(2-24\Cas v_4 l_{1,1}^{(\text{FB}),4}\,
g^2)\, \te \nonumber\\
&&-2 \bigl(\beta_{\lk}^{g^4} \,
g^4+\beta_{\lk}^{h^4}\, h^4\bigr) \, \te^2. \label{bs2}
\end{eqnarray}
(Here, all arguments of the threshold functions which are not
displayed are assumed to be equal to zero; therefore, threshold
functions without any argument are simply numbers which depend on the
details of the regularization). If the scalar field is auxiliary at
the UV scale as in the QCD context, its wave function renormalization
is very small initially, $Z_\phi\ll 1$, so that the dimensionless
renormalized mass is very large, $\epsilon\gg 1$. In this case, scalar
fluctuations are suppressed and the threshold functions depending on
$\epsilon$ vanish; the right-hand side of Eq.\re{bs2} describes a
parabola in the variable $\te$, and we find two positive fixed points,
$0<\te_1^\ast<\te_2^\ast$, where $\te_1^\ast$ is UV attractive but IR
unstable, and $\te_2^\ast$ is an IR stable fixed point (see
Fig.~\ref{figFP}, solid line). It can be shown that $\te_1^\ast$
corresponds to the inverse of the critical coupling of the NJL model,
so that our flow describes a model with strong four-fermion
interaction if we choose UV initial conditions with
$\te_\Lambda<\te_1^\ast$ to the left of the first fixed point (see,
e.g, \cite{Aoki:1996fh} for a detailed analysis of the phase structure
in the abelian case). For
this choice, the system is not in the QCD domain but approaches chiral
symmetry breaking ($\te<0$) in a perturbatively accessible way
(P\xsb$\!\!$). 

In this section, we concentrate on those initial values which release
the system to the right of the first fixed point,
$\te_\Lambda>\te_1^\ast$, i.e., which are weakly coupled in the NJL
language. This will be the range of the QCD universality class. As the
system evolves, it flows towards the second fixed point $\te_2^\ast$,
which then governs the evolution over many scales.  Here, the system
``loses its memory'' of the initial conditions; in particular, it is
of no relevance whether we start with
$\te_1^\ast<\te_\Lambda<\te_2^\ast$ or $\te_\Lambda>\te_2^\ast$. The
evolution towards and in the IR is universally governed by this fixed
point $\te_2^\ast$, which can be shown to be associated with a
fermion-antifermion bound state; e.g., in QED, the properties of the
scalar field at this fixed point correspond to those of positronium
\cite{Gies:2001nw}.

Before we elucidate the fixed-point properties further, let us briefly
mention that its existence can be generalized to the case of a scalar
field describing a fundamental particle in the UV (a Yukawa model with
gauged fermions rather than QCD).  In this case, we have $Z_\phi=1$
and $\epsilon\simeq {\cal O}(1)$ at the UV scale. Now the second term
in Eq.\re{bs2} can become important, in particular for a large
$\phi^4$ coupling $\lambda_\phi$ and/or small $h^2$. When discussing
the RHS of Eq.\re{bs2} for fixed $g$, $h$, $\lambda_\phi$, one should
keep in mind that these couplings may change with $k$. For large
$\lambda_\phi/h^2$, the $\te$ parabola is lowered and the first fixed
point can move to negative values, $\te_1^\ast<0$ (see
Fig.~\ref{figFP}, dashed line). In this case, we can release the
system even in the broken regime at the UV scale, $\te,\epsilon<0$,
but it still evolves towards the bound-state fixed point $\te_2^\ast$.
In comparison with Fig.~\ref{fighit}, this corresponds to initial
values $\bar{m}^2_{\text{c}}<\bar{m}^2_{\Lambda}<0$.  Physically, such
a scenario describes a system involving fundamental scalars, fermions
and gauge fields, where the scalar sector is initially weakly coupled
to the fermions. If we start in the broken regime, scalar fluctuations
will drive the system towards the symmetric regime before the
fermion-gauge-field interactions induce sizable bound-state effects
which can exert an influence on the scalar sector. In this scenario, the
first fixed point $\te_1^\ast<0$ is a measure of the strength of the
initial effective coupling between scalars and fermions. For strong
effective coupling, $\te_\Lambda<\te_1^\ast$, an initial negative
scalar mass of the order of the cutoff, $\mkq\big|_{k=\Lambda}\simeq
-{\cal O}(\Lambda^2)$ will induce a vacuum expectation value and a
fermion mass of the same order, in agreement with naive expectations.
But at weak effective coupling, e.g., $h^2\sim {\cal O}(1)$,
$\lambda_\phi\simeq 100$ and $\te_1^\ast<\te_\Lambda<0$, the system
can still start with an initial negative scalar mass
$\mkq\big|_{k=\Lambda}\simeq -{\cal O}(\Lambda^2)$, but finally run
into the bound-state fixed point. As an important result, the vacuum
expectation value and the fermion mass after symmetry breaking can
easily be orders of magnitude smaller than the UV scale, as exhibited
in Fig.\re{fighit} in the Introduction. We conclude that all systems
with $\te_\Lambda>\te_1^\ast$ belong to the QCD universality class.

Let us now turn to the properties of the system at the bound-state
fixed point. The crucial observation is that not only $\te$ but also
all dimensionless scalar couplings approach fixed points. In the
general case, the fixed-point values depend in a complicated form on
all parameters of the system. However, in the limit $\epsilon\gg1$
(QCD-like), we can find analytic expressions that satisfy the
fixed-point conditions $\pat(\epsilon,h^2,\lambda_\phi)=0$ to leading
order:
\begin{eqnarray}
\epsilon^\ast&\simeq&\frac{2}{|Q_\sigma|}, \label{bs3}\\
(h^\ast)^2&\simeq&\frac{2|\beta_{\lk}^{g^4}| \,g^4}{|Q_\sigma|}
=\frac{12}{|Q_\sigma|} \frac{\Cas (\Nc+2)(\Nc-1)}{\Nc^2}\, v_4
l_{1,2}^{(\text{FB}),4} \, g^4 \nonumber\\
\lambda_\phi^\ast&\simeq&\frac{\Nc\, (h^\ast)^4}{6\, \Cas\,
  g^2}. \nonumber
\end{eqnarray}
From the first equation, we read off that the approximation
$\epsilon\gg 1$ is equivalent to assuming $|Q_\sigma|\ll 1$, which is
roughly fulfilled in our numerical study with our choice of
$Q_\sigma^0=-0.1$. 

The remarkable properties of the IR fixed point become apparent when
considering the renormalized scalar mass, $m^2=\epsilon k^2$. Since
$\epsilon\to \epsilon^\ast$, the scalar mass simply decreases with
the scale $k$, so that it is only {\em natural} to obtain small masses
$m^2\ll \bar{m}^2_{\Lambda}$ for small scale ratios $k\ll
\Lambda$. In other words, even if we start with a scalar mass of the
order of the cutoff, $\mkq\big|_{k=\Lambda}\sim\Lambda^2$, no fine-tuning will be
necessary to obtain small mass values at low-energy scales, as long as
the running is controlled by the bound-state fixed point. 

In order to approach the \xsb regime, the bound-state fixed point has
to be destabilized; otherwise, the system will remain in the symmetric
regime as is the case in QED. In QCD, this destabilization arises from
the increase of the gauge coupling towards the infrared
\cite{Aoki:2000dh}. From the third and last term of Eq.\re{bs2}, it is
obvious that an increasing gauge coupling lifts the $\pat\te$ parabola
(see Fig.~\ref{figFP}, dotted lines). For some value $g_{\text{D}}^2$
of the gauge coupling, the two fixed-points in $\te$ will be
degenerate, so that there is no fixed point at all for all
$g^2>g_{\text{D}}^2$. The beta function $\pat \te$ is then strictly
positive, which drives the system towards the \xsb regime.

In the limit $\epsilon\gg 1$, the critical gauge coupling of
fixed-point degeneracy $g_{\text{D}}^2$ can be computed analytically,
and we find:
\begin{equation}
g_{\text{D}}^2\simeq\frac{16}{3}\pi^2\frac{\Nc}{\Nc-1} \left( \sqrt{
    1+\frac{1}{\Nc+1}}  -1 \right) \simeq \frac{4}{3} \pi^2
    \frac{1}{\Cas}, \label{bs5}
\end{equation}
where we have used linear cutoff functions \cite{Litim:2001up} for
which $l_{1,2}^{(\text{FB}),4}=3/2$.  For instance, for SU(3) we get
$\alpha_{\text{D}} =\frac{g_{\text{D}}^2}{4\pi} \simeq \frac{\pi}{4}$,
which is in the nonperturbative domain, as expected.\footnote{Strictly
speaking, the value of $g_{\text{D}}^2$ is not a physical quantity and
depends on the choice of the cutoff function, i.e., regularization
scheme. This is only natural, since the running of the coupling itself
also depends on the regularization. The scheme dependence, however,
cancels out in physical quantities.} As soon as $g^2$ exceeds
$g_{\text{D}}^2$, the running of the scalar couplings is no longer
protected by the bound-state fixed point. Here all couplings are
expected to run fast, being strongly influenced by the details of the
increase of the gauge coupling. {Of course, owing to strong
coupling, many higher-order operators can acquire large anomalous
dimensions and contribute to the dynamics of the symmetry-breaking
transition. Our truncation should be understood as the minimal
lowest-order approximation in this regime, but gives already a
remarkably consistent (but not necessarily complete) picture.} Once
chiral symmetry is broken, the fermions decouple and the fermionic and
(most of the) scalar flow essentially stops.

The scenario discussed here finally explains why the IR values of the
scalar and fermionic couplings inherit their order of magnitude from
the QCD scale $\Lambda_{\text{QCD}}$ as they should, whereas
particularly the details of the scalar sector at the UV scale are of
no relevance, owing to the fixed-point structure inducing QCD
universality.

\section{Numerical results}
\label{numerics}

In the following, we concentrate on the set of theories that belong to
the QCD universality class. In order to illustrate how universality
arises from the presence of the bound-state fixed point, we initiate
our flows at a GUT-like scale of $\Lambda=10^{15}$GeV, where the gauge
coupling is weak and increases only logarithmically towards the
infrared. Therefore, the bound-state fixed point exists over a wide
range of scales. As discussed before, hardly any dependence on the
specific initial values for the scalar potential and the Yukawa
coupling remains because of the fixed point, as we will demonstrate
quantitatively in the following.

\begin{figure}[t]
\begin{picture}(160,35)
\put(0,0){
\epsfig{figure=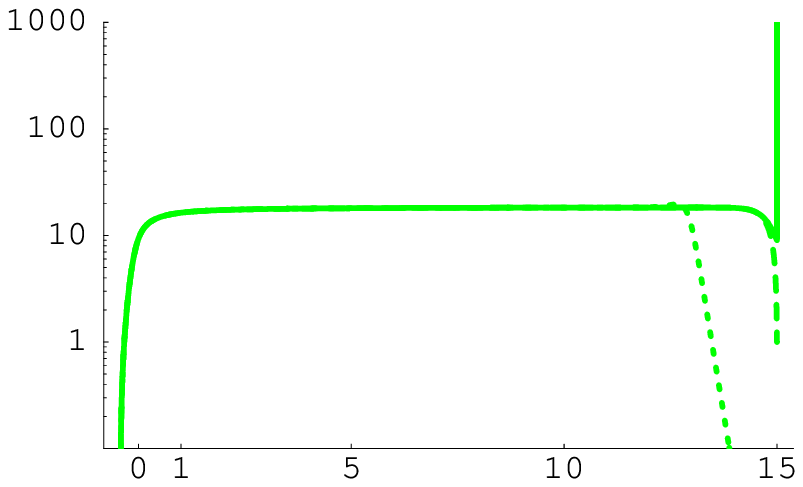,width=5cm}}
\put(25,28){$\epsilon$}
\put(35,-2){$t_{10}$}
\put(48,30){$10^{8}$}
\put(51,8){$1$}
\put(47,4){$10^{-5}$}
\put(74,-7){ $k=10^{t_{10}}$GeV}
\put(55,0){ 
\epsfig{figure=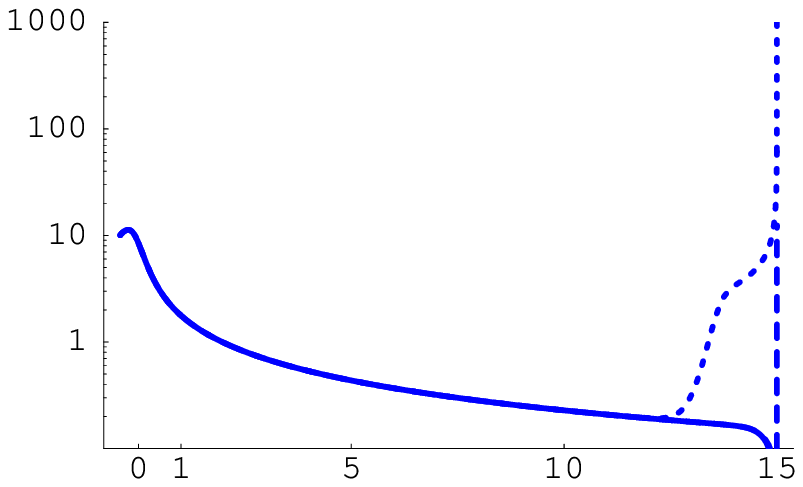,width=5cm}}
\put(80,28){$h$}
\put(90,-2){$t_{10}$} 
\put(103,30){$10^{9/2}$}
\put(106,20){$10^{2}$}
\put(106,3){$10^{-2}$}
\put(110,0){
\epsfig{figure=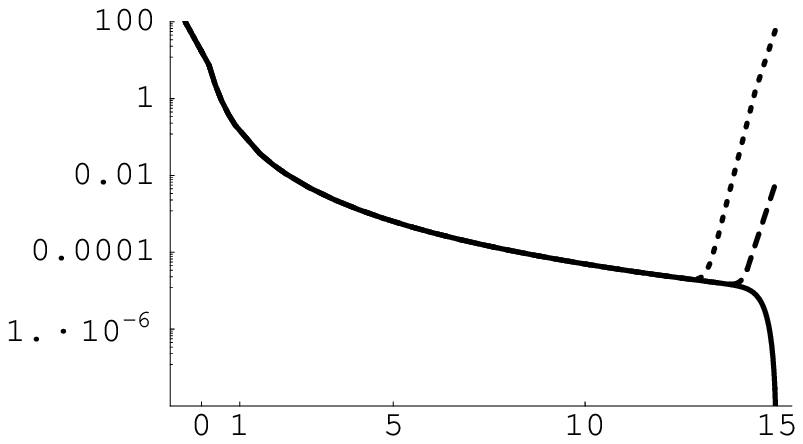,width=5cm}}
\put(135,28){$\lambda_\phi$}
\put(145,0){$t_{10}$}
\put(158,29){$10^{16}$}
\put(160,16){$10^{12}$}
\put(160,5){$0$}
\end{picture} 

\vspace{0.4cm}

\caption{Flow of $\epsilon$, $h$ and $\lambda_\phi$ in the
  symmetric regime according to Eqs.\re{1.13}, \re{1.14}, \re{1.17},
  and \re{1.19}. The solid lines correspond to the reference
  set\re{3.3}, whereas the dotted and dashed lines represent the flows
  for strongly differing initial values as indicated. The
  insensitivity with respect to the choice of initial conditions is
  clearly visible. On the horizontal axis, the exponent $t_{10}$ is
  used for the scale $k=10^{t_{10}}$GeV.}
\label{scalarsym} 
\end{figure}

For illustrative purposes, we concentrate here on QCD-like scenarios
where the scalar is auxiliary at the UV scale, and explore this
parameter space using the natural choice given by Eq.\re{1.5} as a
reference; to be precise, we use the reference set,
\begin{eqnarray}
&&\mkq\big|_{k=\Lambda}=\Lambda^2,\quad
\bar{\lambda}_{\phi}\big|_{k=\Lambda}=0, 
\quad\Zphi\big|_{k=\Lambda}=10^{-8},\quad
\bar{h}^2\big|_{k=\Lambda}=10^{-12}, \nonumber\\
\Leftrightarrow&&\epsilon\big|_\Lambda=10^8,\quad
\lambda_\phi|_\Lambda=0,\quad h|_\Lambda=10^{-2},\label{3.3} 
\end{eqnarray}
in our numerical studies. In all computations, we use linear cutoff
functions proposed in \cite{Litim:2001up} for which the threshold
functions can be determined analytically (see App.~\ref{threshold}).
We plot the flows of the renormalized dimensionless couplings
$\epsilon$, $h$ and $\lambda_\phi$ in Fig.~\ref{scalarsym} for the
symmetric regime. The reference set \re{3.3} is depicted as solid
lines, whereas the dashed and dotted lines correspond to initial
values which deviate from the reference set\re{3.3} by many orders of
magnitude for the corresponding couplings.

As long as we start in the range of attraction of the bound-state
fixed point, we can obviously vary the initial values for the scalar
couplings over many orders of magnitude without any appreciable
effect. The system quickly approaches the bound-state fixed point,
where the initial values of the couplings become unimportant. In
particular, the scalar mass, which is allowed to be of the order of
the cutoff or even much larger at $k=\Lambda$, runs to small values
$\sim k$ while the system is governed by the bound-state fixed point.
No fine-tuning is necessary for this.\footnote{As a fairly weak
  condition, we only have to ensure that the initial scalar couplings
  are such that no strong four-fermion interaction is implicitly
  induced by the initial values; in this case, the system starts to
  the left of the first fixed point $\te_1^\ast$ and rather resembles
  a nonabelian gauged NJL model with P\xsb.} Let us stress once more
that these features of universality are not restricted to the
reference set\re{3.3} and the variations thereof. They can also be
found in Yukawa models with a fundamental scalar
($Z_\phi|_{k=\Lambda}=1$) and even if we start in the broken regime at
the UV scale (see Fig.~\ref{fighit}).

At the bound-state fixed point, the couplings are modulated only by
the logarithmically slow increase of the gauge coupling. Incidentally,
the modulation of $\te=\epsilon/h^2$ is completely carried by $h$,
whereas $\epsilon$ stays fixed. This agrees with our analytical
fixed-point values found in Eq.\re{bs3}. A rapid change for the couplings in
Fig.~\ref{scalarsym} is visible after $g^2$ exceeds $g_{\text{D}}^2$
and the bound-state fixed point has disappeared ($t_{10}\lesssim
1$).

The behavior of the system changes rapidly after the gauge coupling
has grown large. For $g^2>g_{\text{D}}^2$, the bound-state fixed point
vanishes and all couplings start to run fast. The system necessarily
runs into the \xsb regime where the scalars develop a vacuum
expectation value and the fermions acquire a mass
\begin{equation}
m_{\text{f}}^2=\lim_{k\to 0} k^2\, \kappa h^2\equiv (h
\sigma_{\text{R}})^2, \label{3.1}
\end{equation}
where $\sigma_{\text{R}}=\lim_{k\to 0}\sqrt{\Zphi\, \rho_0}$ denotes
the renormalized expectation value of the scalar field. 

This leads to a decoupling of the fermions, and, consequently,
fermion-boson translation is ``switched off''. Also the flow of the
Yukawa coupling stops, the scalar and fermion anomalous dimensions
approach zero, and $\kappa$ runs according to its trivial mass
scaling, $\kappa\sim 1/k^2$, so that $m_{\text{f}}$ approaches a
constant value.

Whereas the qualitative picture is rather independent of the details
of the running gauge coupling, quantitative results are highly
sensitive to the flow of the gauge sector. This is because a finite
amount of ``RG time'' passes from the disappearance of the bound-state
fixed point to the transition into the \xsb regime. In between, the
running of the gauge coupling exerts a strong influence on all other
couplings which are no longer protected by any fixed point. A purely
perturbative running of the gauge coupling turns out to be
insufficient for the present purpose, since the (unphysical) Landau pole
destabilizes the system in the infrared. 

For definiteness, let us consider a running coupling governed by the
beta function
\begin{eqnarray}
\pat g^2=\beta_{g^2}=\etaF g^2&=&-2\left(b_0\, \frac{g^4}{16\pi^2}+ b_1\,
\frac{g^6}{(16\pi^2)^2} \right) \left[ 1-\exp\left(\frac{1}{\alpha_\ast}
-\frac{1}{\frac{g^2}{4\pi}} \right)\right]^s, \label{3.2}\\
b_0&=&\frac{11}{3} \Nc -\frac{2}{3} N_{\text{f}},\quad 
b_1=\frac{34}{3} \Nc^2- \frac{10}{3}\Nc N_{\text{f}} 
    -2\Cas N_{\text{f}} \nonumber
\end{eqnarray}
for our numerical studies.  In the UV, this beta function exhibits an
accurate two-loop perturbative behavior, whereas the coupling runs to
a fixed point $\alpha_{\text{s}}\equiv g^2/(4\pi)\to \alpha_\ast$ in
the IR for $k\to 0$. In the first place, the infrared fixed point is
convenient for numerical purposes, since it does not lead to
artificial IR instabilities. Moreover, an infrared fixed point for a
mass-scale-dependent running coupling is compatible with the expected
mass gap in Yang-Mills theory. Below this mass gap, all gauge field
fluctuations decouple from the flow and can no longer drive the flow
of the coupling. Different beta functions with and without infrared
fixed points are studied in Appendix \ref{beta}. It turns out that,
though the infrared properties such as the constituent quark mass
depend quantitatively on the choice of the beta function as expected,
the universal features discussed in the following remain
untouched. This underlines our observation that the detailed
understanding of the flow for the region of strong gauge coupling is
not essential for the overall picture.

In combination with Eq.\re{3.2}, the system of flow equations is now
closed and provides us with an answer for the (truncated) quantum
effective action, once we specify all parameters and initial values.
We have investigated SU($\Nc=3$) gauge theory with initial value
$g(\Lambda)$ chosen such that $\alpha_{\text{s}}$ acquires its
physical value at the $Z$-boson mass, $\alpha_{\text{s}}(M_Z)\simeq
0.117$. We work in the {background} Landau gauge, $\xi=0$, which
is known to be a fixed point of the renormalization flow in the gauge
sector \cite{Ellwanger:1995qf},\cite{Litim:1998qi}. If we had an exact
flow equation at our disposal this choice would fix the system
completely.

In our truncation, however, we have the parameter $Q_\sigma^0$, in
addition to the Yang-Mills beta function, which characterizes our
ignorance of the exact flow. The quantity $Q_\sigma^0$ measuring the
momentum suppression of the four-fermion interaction will be set to
$Q_\sigma^0=-0.1$, in agreement with our considerations given above.
It turns out that the infrared properties of the system are only
weakly dependent on this parameter and on $\xi$ (see below), which
substantiates our truncation.  Furthermore, we choose $\alpha_\ast$ to
be of order 1, but not too close to $g_{\text{D}}^2/(4\pi)$ in order
to avoid pathologies: $\alpha_\ast=2.5$.

\begin{figure}[t]
\begin{picture}(160,45)
\put(0,0){
\epsfig{figure=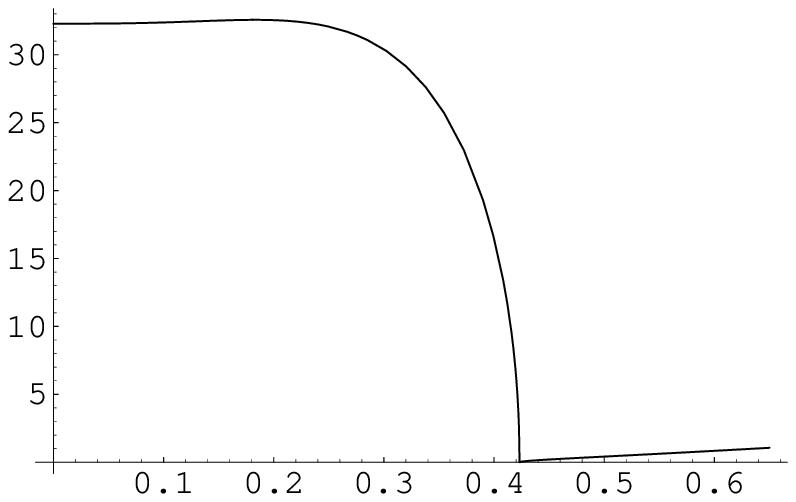,width=7.5cm,height=4cm}}
\put(0,40){MeV}
\put(68,6){$m$}
\put(30,40){$\sigma_{\text{R}}$}
\put(65,-4){$k$/GeV}
\put(84,40){MeV}
\put(84,0){ 
\epsfig{figure=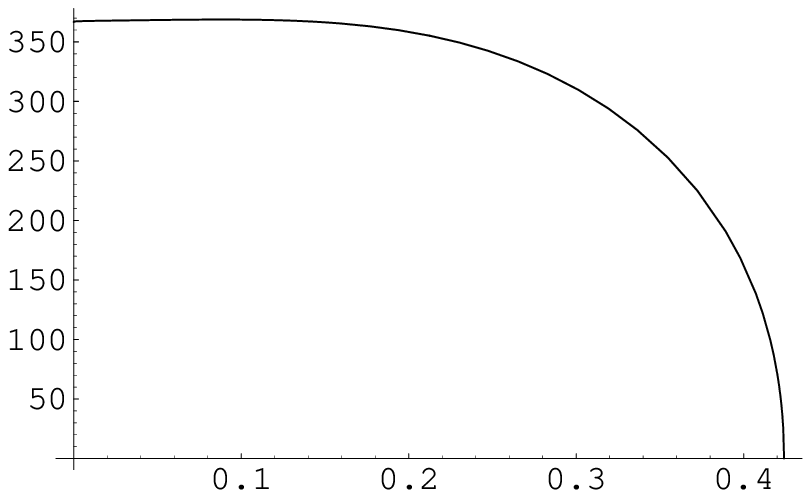,width=7.5cm,height=4cm}}
\put(128,38){$m_{\text{f}}$}
\put(146,-4){$k$/GeV}
\end{picture} 
\caption{Flow of the scalar mass $m$, the scalar VEV
  $\sigma_{\text{R}}$, and the constituent quark mass $m_{\text{f}}$
  close to and in the \xsb regime, using the reference set\re{3.3}.
  For the particular choice for the running of the gauge coupling
  according to Eq.\re{3.2} with $\alpha_\ast=2.5$, the transition
  occurs at $k_{\text{\xsb}}\simeq423$MeV.}
\label{VEV} 
\end{figure}

\begin{figure}[t]
\begin{picture}(160,85)
\put(0,45){
\epsfig{figure=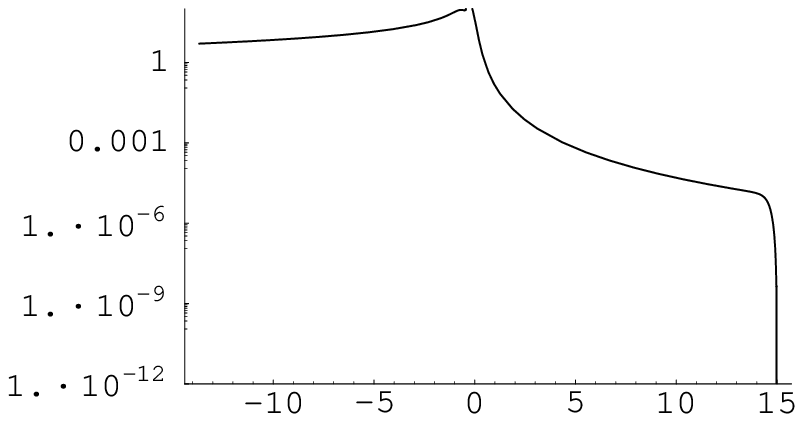,width=7.5cm,height=4cm}}
\put(60,80){$\lambda_\phi$}
\put(67,45){$t_{10}$}
\put(84,45){ 
\epsfig{figure=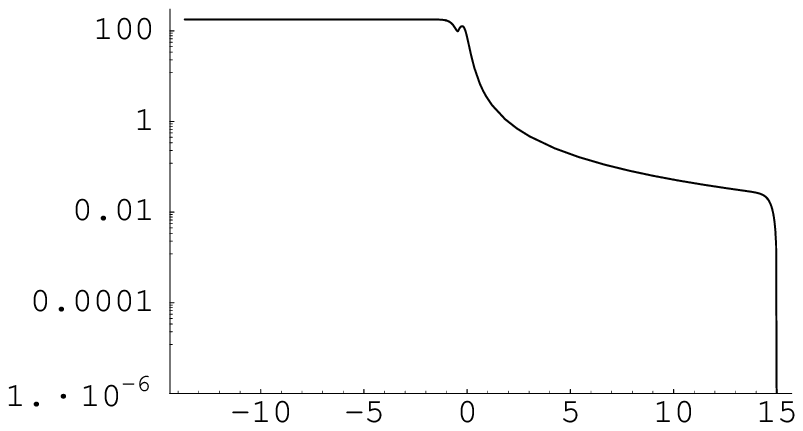,width=7.5cm,height=4cm}}
\put(138,80){$h^2$}
\put(149,45){$t_{10}$}
\put(0,0){
\epsfig{figure=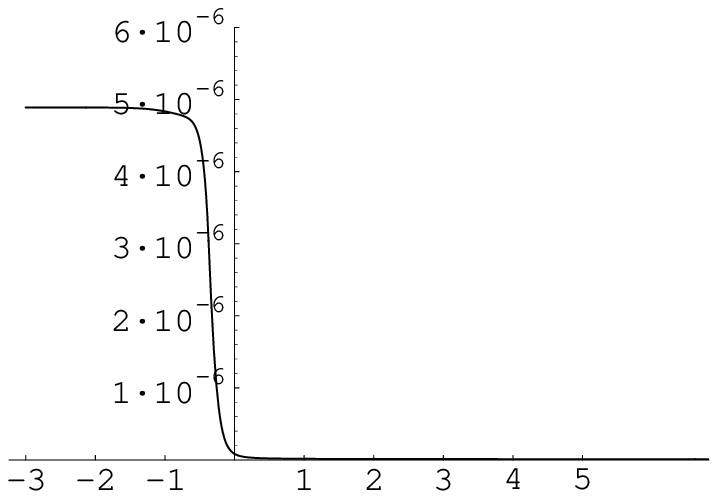,width=7.5cm,height=4cm}}
\put(6,34){$\Zphi$}
\put(62,-1){$t_{10}$} 
\put(84,0){  
\epsfig{figure=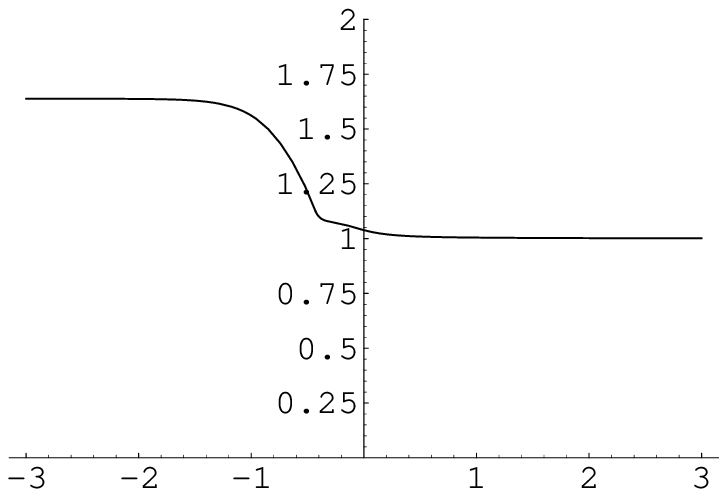,width=7.5cm,height=4cm}}
\put(135,25){$\Zy$}
\put(148,-1){$t_{10}$}
\put(73,-3){ $k=10^{t_{10}}$GeV}
\end{picture} 
\caption{Flow of $\lambda_\phi$, $h^2$, and the wave function
  renormalizations $\Zphi$ and $\Zy$ over the complete range of scales
  for the reference set\re{3.3}. The rapid change of all couplings
  near $t_{10}=\log_{10} k_{\text{\xsb}}/\Lambda \simeq-0.5$ is
  visible. Whereas $h^2$, $\Zphi$ and $\Zy$ approach fixed points in
  the deep infrared owing to decoupling, $\lambda_\phi$ decreases
  logarithmically owing to a massless ``eta'' in absence of the axial
  anomaly. }
\label{collect} 
\end{figure}

For this concrete scenario, the transition to the \xsb regime occurs at
$k_{\text{\xsb}}\simeq 423$MeV. The renormalized scalar mass slightly
above $k_{\text{\xsb}}$ and the VEV of the scalar field below
$k_{\text{\xsb}}$ are depicted in Fig.~\ref{VEV} (left panel).
According to Eq.\re{3.1}, we find a constituent quark mass of
$m_{\text{f}}\simeq 371$MeV as shown in Fig.~\ref{VEV} (right
panel). Of course, these numbers depend strongly on the details of the
Yang-Mills beta function for strong coupling $\alpha_{\text{s}}\sim
1$; various other examples are discussed in Appendix \ref{beta}. 
Finally, the running of $\lambda_\phi$, $h^2$ and the scalar and
fermionic wave function renormalizations is collected in
Fig.~\ref{collect}.

Focusing on low-energy QCD-like aspects of our truncated system, it is
also remarkable that (apart form the scalar couplings) the choice of
$Q_\sigma^0$ has little effect on infrared properties of the system:
varying $Q_\sigma^0$ between $-0.5\dots0.001$ changes
$k_{\text{\xsb}}$ or $m_{\text{f}}$ only at the level of less than
10\%.  This is reassuring and in contrast to the strong
$Q_\sigma^0$-dependence of the bound-state fixed-point values of
$\epsilon_\ast$ and $h_\ast$. The variations of the infrared
properties are similarly small for changes in the gauge parameter in
the interval $\xi=0\dots 2$.

To summarize, a large class of QCD-like theories including a scalar
degree of freedom belong to the QCD universality class owing to an
attractive infrared fixed point present for weak gauge coupling. Even
before the gauge coupling becomes strong, all theories in this
universality class are indistinguishable at low energies. They exhibit
an identical approach to \xsb which is triggered and quantitatively
determined by the increase of the gauge coupling.

\section{Instanton-mediated interactions, axial anomaly\\ and the
  fate of the eta boson}
\label{anomaly}

Up to now, we have considered only that part of the model which has a
global $\UA$ symmetry corresponding to simultaneous axial phase
rotations of the scalars and fermions. In QCD, this symmetry is
anomalously broken by the presence of gauge-field configurations of
nontrivial topology. For instance, instantons induce fermion
interactions which break this symmetry. In an
instanton--anti-instanton background, the $\Nf=1$ interaction is
mass-like and can be expressed as \cite{'tHooft:fv}
\begin{eqnarray}
{\cal L}_{\text{I+A}}&=&\int_0^{\bar{f}_{\text{c}}(k,m_{\text{f}})}
\frac{d\rhov}{\rhov^5}\, 
d_0^{\Nc}(\rhov)\, C_{\text{E}}(\Nc)\, (2\pi^2\rhov^3)\, 
\left(  \frac{\alpha(1/\rho)} {\alpha(\bar{\mu})}\right)^{-4/b_0}
(\ybr\yl-\ybl\yr), \label{4.1}\\
&&d_0^{\Nc}(\rhov):=\frac{4.6\, \E^{-1.68\Nc}}{\pi^2 (\Nc-1)!(\Nc-2)!}
\left( \frac{2\pi}{\alphas(1/\rhov)}\right)^{2\Nc}
\E^{-\frac{2\pi}{\alphas(1/\rhov)}}, \nonumber
\end{eqnarray}
where $C_{\text{E}}(\Nc)$ is a color factor that arises from averaging
over all possible embeddings of $SU(2)$ into $SU(\Nc)$, e.g.,
$C_{\text{E}}(2)=1$, $C_{\text{E}}(3)=2/3$, and $\bar{\mu}$=1GeV is
the renormalization scale for the fermion fields.  Note that we
introduced an IR cutoff function $\bar{f}_{\text{c}}(k,m_{\text{f}})$
in the upper bound of the instanton radius $\rhov$ integration. This
function should cut off the contribution from all modes with momenta
either below $k$ or the generated fermion mass $m_{\text{f}}$, and
thereby implements the renormalization group formulation of this
interaction in a simple manner. The $\rhov$ integration is UV finite
for $\rhov\to 0$ owing to asymptotic freedom, and the infrared
($\rhov\to\infty$) is controlled by the cutoff $\bar{f}_{\text{c}}$
and by the increase of the coupling.
 
{In the following, we intend to include this interaction as it is,
being an example for a $\UA$ violating term. Contrary to standard
instanton based models \cite{ilm}, we do not employ further
information about, e.g., average instanton sizes and separations or
other assumptions about the vacuum state of the gauge field. For
this,} we note that Eq.\re{4.1} already corresponds to an integrated
flow, ${\cal L}_{\text{I+A}}=\mIA\, (\ybr\yl-\ybl\yr)$, where the flow
of the induced mass $\mIA$ is given by\footnote{A more rigorous
treatment of anomalous $\UA$ breaking within the flow equation
formalism has been suggested in \cite{Pawlowski:1996ch}.}
\begin{equation}
\pat \mIA=2\pi^2  \Zy \left[d_0^{\Nc}(\rhov)\,
  C_{\text{E}}(\Nc)\,\left(  \frac{\alpha(1/\rho)}
    {\alpha(\bar{\mu})}\right)^{-4/b_0}
  \right]_{\rhov=\bar{f}_{\text{c}}(k,m_{\text{f}})} \pat
  \bar{f}_{\text{c}}(k,m_{\text{f}}), \label{4.2}
\end{equation}
with the initial condition $\mIA(k=\Lambda\to\infty)\to 0$. For
consistency, we also included here the fermion wave function
renormalization $\Zy$, which was not taken into account in Eq.\re{4.1}
as derived in \cite{'tHooft:fv}. Since $\bar{f}_{\text{c}}$ has mass
dimension -1, an appropriate choice is given by
\begin{equation}
\bar{f}_{\text{c}}(k,m_{\text{f}})=\frac{1}{k}\, f_{\text{c}}(\kappa
h^2),\quad \text{with}\,\,f_{\text{c}}(0)=1,\,\,f_{\text{c}}(\kappa
h^2)\bigl|_{\kappa h^2\to \infty}\to \frac{1}{\sqrt{\kappa h^2}}
\label{4.2a}, 
\end{equation}
such that $\bar{f}_{\text{c}}(0,m_{\text{f}})=1/m_{\text{f}}$. For
our numerical solutions, we will use $f_{\text{c}}(x)=(1+x)^{-1/2}$
for simplicity. With these definitions, we can rewrite Eq.~\re{4.2} as
\begin{equation}
\pat\mIA=-2\pi^2  \Zy \, \frac{k}{f_{\text{c}}}\,
  d_0^{\Nc}(f_{\text{c}}/k)\,  C_{\text{E}}(\Nc)\,
 \left(  \frac{\alpha(k/f_{\text{c}})}
    {\alpha(\bar{\mu})}\right)^{-4/b_0}
   \left(1+ \frac{(-f_{\text{c}}')}{f_{\text{c}}}\, \pat
   (\kappa h^2) \right), \label{4.2b}
\end{equation}
where $f_{\text{c}}=f_{\text{c}}(\kappa h^2)$, and the prime denotes a
derivative.

Now we could repeat the calculation of the flow equations of
Sect.\ref{setting} including this fermion mass term in the
propagator. In this way, however, we would induce a number of $\UA$
noninvariant fermion-fermion and fermion-scalar couplings which 
complicate the calculation unnecessarily. Instead, we propose a
generalization of the field transformation\re{d13ba} which serves to
translate the instanton-induced interaction into the scalar sector:
\begin{eqnarray}
\pat\phi_k(q)&=&-(\ybl\yr)(q)\, \pat\alpha_k(q)+ \phi_k(q)\,
\pat\beta_k(q)+\pat\gamma_k +(\phid_k\phi_k)\phi_k\, \pat\delta_k,
\nonumber\\ 
\pat\phid_k(q)&=&(\ybr\yl)(-q)\, \pat\alpha_k(q)+ \phid_k(q)\,
\pat\beta_k(q)+\pat\gamma_k +(\phid_k\phi_k)\phid_k\, \pat\delta_k
\label{4.3},
\end{eqnarray}
with additional a priori arbitrary functions $\gamma_k$ and
$\delta_k$, whereas $\alpha_k$ and $\beta_k$ are those of
Sect.~\ref{setting}. The term $\sim \pat \gamma_k$ corresponds to a
$\UA$ violating shift of the scalar field which can compensate for
the instanton-induced fermion mass. The flow of $\mIA$ is now
given by
\begin{equation}
\pat\mIA=\pat\mIA\big|_{\phi_k} +\hk\,
\pat\gamma_k-\frac{1}{2}\nub \, \pat\alpha_k,\label{4.4}
\end{equation}
where the second and third terms arise from the transformation of the
Yukawa interaction and the last term in Eq.\re{1.3}, respectively. Now
we can determine $\gamma_k$ such that $\pat\mIA=0$ holds on all
scales. In this way, the instanton interaction does not affect $\mIA$
(which vanishes on all scales), but is translated into the scalar
sector and contributes to the running of $\nub$. In the point-like
limit ($q^2=0$), we find:
\begin{equation}
\pat\nub=-2\mkq\, \pat\gamma_k +\nub\, \pat\beta_k. \label{4.5}
\end{equation}
Introducing the dimensionless renormalized quantity
\begin{equation}
\nu=\frac{\nub}{\Zphi^{1/2} k^3}, \quad \Rightarrow\quad
\nu_{\text{R}}=k^3\, \nu, \label{4.6}
\end{equation}
where $\nu_{\text{R}}$ denotes the renormalized (dimensionful) value,
we finally arrive at
\begin{eqnarray}
\pat\nu&=&-\left(3-\frac{\eta_\phi}{2} \right)\nu-4\pi^2
\frac{\epsilon}{h}\, d_0^{\Nc}(f_{\text{c}}/k)\,
C_{\text{E}}(\Nc)\,\left(  \frac{\alpha(k/f_{\text{c}})}
    {\alpha(\bar{\mu})}\right)^{-4/b_0} \frac{1}{f_{\text{c}}}
   \left(1+ \frac{(-f_{\text{c}}')}{f_{\text{c}}}\, \pat
   (\kappa h^2) \right) \nonumber\\
&&+\frac{\nu}{h^2} \bigl(1+(1+\epsilon)^2Q_\sigma\bigr)
\bigl(\beta_{\lk}^{g^4} \, g^4+\beta_{\lk}^{h^4}\, 
   h^4\bigr),\label{4.7}
\end{eqnarray}
which describes the running of the axial anomaly in the instanton
approximation. 

The shift $\sim \pat \gamma_k$ induces another $\UA$ violating term
$(\phid\phi)(\phid+\phi)$ via the transformation of the
$\lambda_\phi(\phid\phi)^2$ term. This can be cancelled by an
appropriate choice of the last transformation function $\delta_k$ in
Eq.\re{4.3}, which has to satisfy
\begin{equation}
\lf\, \pat\gamma_k -\frac{1}{2} \nub\, \pat\delta_k =0. \label{4.8}
\end{equation}
Finally, the terms $\sim \delta_k$ in Eq.\re{4.3} influence the
running of $\lambda_\phi$ via the transformation of the scalar mass
term. The modified flow equation for $\lambda_\phi$ reads:
\begin{eqnarray}
\pat\lambda_\phi&=&\pat\lambda_\phi\big|_{\phi_k}
+4\frac{\lambda_\phi}{h^2} 
\bigl(1+2\epsilon+(1+\epsilon)^2Q_\sigma\bigr)
\bigl(\beta_{\lk}^{g^4} \, g^4+\beta_{\lk}^{h^4}\, h^4\bigr) \nonumber\\
&&+16\pi^2 \frac{\epsilon\lambda_\phi}{\nu h} \,
d_0^{\Nc}(f_{\text{c}}/k)\, C_{\text{E}}(\Nc)\, \left(
  \frac{\alpha(k/f_{\text{c}})} {\alpha(\bar{\mu})}\right)^{-4/b_0}
  \frac{1}{f_{\text{c}}} \left(1+
    \frac{(-f_{\text{c}}')}{f_{\text{c}}}\, \pat (\kappa h^2)
  \right).\label{4.9} 
\end{eqnarray}
These equations are valid in the symmetric regime with similar
equations for the \xsb regime displayed in appendix \ref{nuflow}.

Strictly speaking, the system is never in the symmetric regime, since
chiral symmetry is always broken implicitly by a nonzero $\nu$ term
which induces a nonzero VEV $\sigma_0$ for the scalar field.  For
instance, rotating the VEV into the real component,
$\phi=\sigma_0=\phid$, $\sigma_0=\sqrt{\rho_0}$, the location of the
minimum obeys
\begin{equation}
0=U'(\rho_0)=\mkq+\lf\rho_0-\frac{\nub}{2\sqrt{\rho_0}} \quad
\Rightarrow\quad 
0=\epsilon+\kappa\lambda_\phi-\frac{1}{2}\frac{\nu}{\sqrt{\kappa}}.
\label{4.10}
\end{equation}
Obviously, $\kappa=0$ is not allowed if $\nu\neq0$, owing to the
linear term in $\phi$ in Eq.\re{1.3}. The running of the minimum can
be inferred from
\begin{equation}
0=\pat U'(\rho_0)\big|_\rho=U''(\rho_0)\, \pat\rho_0 +\pat
U'(\rho_0)\big|_{\rho_0} \quad\Rightarrow\quad
\pat\rho_0=-\frac{1}{\lf+\frac{\nub}{4}\rho_0^{-3/2}}\, \pat
U'(\rho_0)\big|_{\rho_0}.\label{4.11}
\end{equation}
Since the instanton-induced terms are exponentially small for the
major part of the flow, the minimum of the potential is actually very
close to zero, and the equations for the symmetric regime of
Sect.~\ref{setting} can be used up to tiny corrections. The solution
of the flow equations is numerically difficult with an exponentially
small $\kappa$ in the broken regime. Therefore, we decide to solve the
flow equations for large enough $k$ in the symmetric-regime formulation.
In this regime, $\nu$ evolves according to Eq.\re{4.7} with only a
subdominant coupling to the other flow equations via Eq.\re{4.9}. Then
we switch to the broken-regime description at that scale where the
instanton-induced fermion mass $m_{\text{f}}$ is of the order of a few
MeV; this procedure induces an error only at the per-mille level and
turns out to be insensitive to the details of the switching scale.

We have analyzed the flow equations including the instanton-mediated
interaction numerically and used the reference set of initial
conditions as defined in Sect.~\ref{numerics} (see Eq.\re{3.3}) for a
direct comparison. As expected, most properties of the system are
unaffected by the instantons, while the system is governed by the
bound-state fixed point. Here the instanton-induced effects are
exponentially suppressed, since the coupling is small. In particular,
the running of the scalar mass $\epsilon$ and the Yukawa coupling are
identical to the ones displayed in Fig.~\ref{scalarsym}, and the
universality properties discussed in Sect.~\ref{numerics} remain
unaffected. 

\begin{figure}[t]
\begin{picture}(160,45)
\put(33,41){$(\text{GeV})^3$}
\put(0,0){
\epsfig{figure=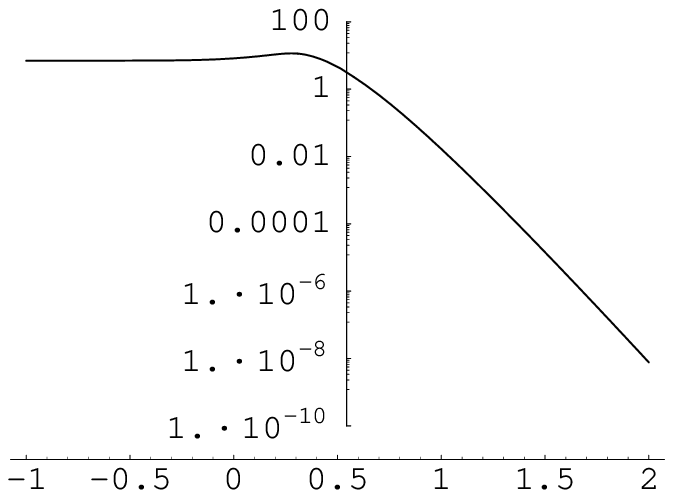,width=7.5cm,height=4cm}}
\put(52,34){$\nu_{\text{R}}$}
\put(63,-2){$t_{10}$}
\put(20,-6){ $k=10^{t_{10}}$GeV}
\put(84,-4){
\epsfig{figure=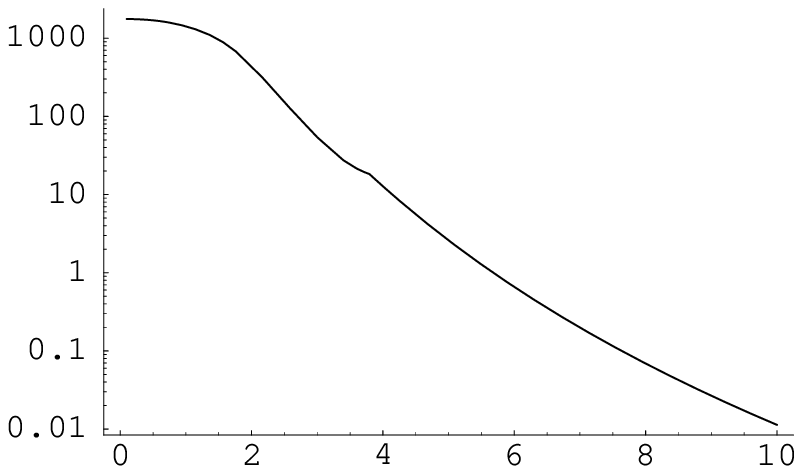,width=8cm,height=5cm}}
\put(84,45){$\text{MeV}$}
\put(120,40){$m_{\text{f}}$}
\put(151,-5){$k/$GeV}
\end{picture}
 
\vspace{0.4cm}

\caption{Left panel: axial anomaly $\nu_{\text{R}}$ in the vicinity of
  the scale of fermion decoupling. Right panel: instanton-induced
  fermion mass. Both plots refer to the reference set\re{3.3} and the
  particular choice for the running of the gauge coupling according to
  Eq.\re{3.2} with $\alpha_\ast=2.5$. Comparison with Fig.~\ref{VEV}
  shows that the fermion mass is dominated by instanton effects.}
\label{instplot} 
\end{figure}
 
The renormalized axial anomaly $\nu_{\text{R}}$ is plotted in
Fig.~\ref{instplot} (left panel). It remains exponentially small for
a large part of the flow and becomes of order $(\text{GeV})^3$ and
larger only in the strong-gauge-coupling regime. Here, however, it
contributes strongly to the VEV of the scalar field and consequently
to the constituent quark mass which leads to the decoupling of the
fermions.

We observe a rather smooth onset of fermion-mass
generation. Furthermore, the constituent quark mass is strongly
enhanced by the instanton interactions. For the reference set, we find
$m_{\text{f}}= 1765$MeV in the infrared limit $k\to 0$.
Again, this number depends strongly on the precise choice of the
running gauge coupling in the infrared, and a number of other
possibilities including instanton effects is listed in Appendix
\ref{beta}.

Let us finally discuss the fate of the ``would-be'' Goldstone boson,
which we may call the eta boson in the style of real QCD. Neglecting
the axial anomaly, this boson arises from spontaneous breakdown of the
global $\UA$ as a true massless Goldstone boson; its effects on the
scalar sector even after \xsb are visible in the logarithmic running
of the scalar $\phi^4$ coupling $\lambda_\phi$ as can be seen in
Fig.~\ref{collect}. The $\UA$ anomaly, however, generates a mass of
the eta boson. In the present formulation, the $\UA$ anomaly occurs as
the $\bar{\nu}$ term in the scalar potential\re{1.3}. Its contribution
to the renormalized eta mass can be computed as
\begin{equation}
m_{\eta}^2=\frac{\nu_{\text{R}}}{2
  \sigma_{\text{R}}}. \label{6.1}
\end{equation}
Within the above-given framework of instanton-mediated interaction, we
find for the eta boson mass in the QCD universality class a value of
$m_{\eta}\simeq 4440$MeV. Of course, this value also strongly depends
on the choice of the running of the gauge coupling and should be used
only for comparison with other masses computed for the same running
gauge coupling. In particular, we find roughly the ratio
$m_\eta/m_{\text{f}}\simeq 3$. This scenario giving rise to a heavy
mass of a would-be Goldstone boson is familiar from three-flavor QCD.

By contrast, the fate of the eta boson is more spectacular if we
go beyond the border of the QCD domain to that of P\xsb$\!\!$,
corresponding to a choice of $\bar{m}^2_\Lambda<\bar{m}^2_{\text{c}}$
in Fig.~\ref{fighit} or $\te_\Lambda<\te_1^\ast$ in Fig.~\ref{figFP}.
Here, the VEV of the scalar field is generically of the order of the
cutoff $\Lambda=10^{15}$GeV.  At the same time, the fermions rapidly
become massive and decouple from the flow only a little below
$\Lambda$. As a consequence, instanton contributions or other
long-distance topological properties have little effect on the fermion
sector and thus the axial anomaly exerts hardly any influence on the
scalars. As a result, the contributions to the eta mass are strongly
suppressed -- powerlike in the denominator and exponentially in the
numerator. For instance, for the set of initial parameters
corresponding to Fig.~\ref{fighit} (right panel) with
$\bar{m}^2_\Lambda$ slightly below $\bar{m}^2_{\text{c}}$, we find an
extremely small eta mass, $m_{\eta}\simeq 2\cdot10^{-30}$eV. For
smaller $\bar{m}^2_\Lambda$, the eta mass decreases even further, and
larger eta masses require a tremendous fine-tuning of
$\bar{m}^2_\Lambda$ close to $\bar{m}_{\text{c}}^2$.

In this scenario beyond the QCD universality class, we have thus found
a mechanism to generate extremely small masses without any
fine-tuning. From another perspective, this mechanism exploits the
fundamentally different RG properties of scalars and chiral gauge
theories. For systems in the universality class of P\xsb$\!$, the \xsb
scale of the scalar sector is generically of the order of the UV
scale, whereas the nonperturbative scale of the gauge sector can be
much smaller. Now the mass of the would-be Goldstone boson is
generated by the nonperturbative sector of the gauge theory which is
exponentially suppressed at the UV scale. This interplay finally leads
to the generation of the extremely small mass.

\section{Conclusions}

In this work, we studied a class of theories involving one-flavor
massless QCD and a chiral color-singlet scalar field. Our model is
parametrized by the gauge coupling and a number of scalar couplings.
In this framework, we identified the QCD universality class of
theories which share the same physics at low energies, namely
spontaneous breaking of chiral symmetry triggered by the strongly
interacting gauge sector at the QCD scale. As a remarkable result, the
QCD universality class contains theories with fundamental scalars
where the microscopic scalar potential has its minimum at nonzero
field ($\bar{m}_{\Lambda}^2>\bar{m}_{\text{c}}^2\sim -{\cal
  O}(\Lambda^2)$). {For these the theories, the scalar
fluctuations drive the system first into the symmetric regime with a large
positive scalar mass, and the remaining flow is governed by the QCD
sector. We checked explicitly that this is in accord with perturbative
expectations for weak couplings (cf. Fig.~\ref{fighit}, right panel).} 

The mechanism that establishes QCD universality is the occurrence of
an infrared attractive bound-state fixed point in the scalar couplings
which persists over a wide range of scales as long as the gauge
coupling is weak. At this fixed point, the scalar field exhibits
quark-antiquark bound-state behavior and the RG running of the scalar
couplings is governed by the RG behavior of QCD. All memory of the
scalar initial conditions is lost by the system. As a remarkable
consequence, the scalar mass is not a relevant operator at this fixed
point. For increasing gauge coupling, the bound-state fixed point is
destabilized and the system runs towards the \xsb regime. Here the
role of the scalar field changes and it can characterize (quark)
condensates and (mesonic) excitations on top of the
condensate. {At strong coupling, the simple overall picture of
\xsb arising from our truncation can, of course, be modified
quantitatively as well as qualitatively by the influence of
higher-order operators. In particular, mixed non-minimal fermion-gluon
and scalar-gluon operators might add new features to \xsb by providing
a coupling to the nontrivial gluonic vacuum structure.}

Beyond the QCD universality class, we find the class of theories
exhibiting perturbative spontaneous chiral symmetry breaking
(P\xsb$\!\!$). In this class, the system is mainly driven by the
scalar sector, and IR properties such as condensates and generated
fermion masses depend strongly on the initial scalar parameters. The
gauge sector exerts hardly any influence on the fermions in this class
unless the scalar parameters are fine-tuned to a high precision. In
the deep IR, pure gluodynamics without dynamical quarks remains. The
flow of the scalar couplings is never in the attractive domain of the
bound-state fixed point, but is governed by a fundamental-particle
fixed point. Small deviations from this fixed point have an infrared
unstable component which corresponds to the RG relevant scalar-mass
operator.

In both universality classes, we found interesting implications. Our
setup of the QCD universality class admits a resolution of an old
puzzle: whereas QCD has no fine-tuning problem and is completely
determined by fixing the coupling at a certain scale, low-energy QCD
models based on NJL-type fermion self-interactions depend strongly on
additional parameters such as an intrinsic UV cutoff. In the context
of partial bosonization, this cutoff-dependence corresponds to a
strong dependence of IR observables on the bosonization scale (or the
value of the scalar mass at this scale). In our approach with
scale-dependent field transformations, partial bosonization occurs at
all scales, and no artificial dependence on unphysical scales is
introduced. In our truncation, QCD flows continuously from a high
scale with quarks and gluons as the relevant degrees of freedom to
intermediate scales with quarks, gluons and quark bound states and
further to low scales with constituent quarks, condensates and mesons.

In the P\xsb universality class with one fermion flavor, we identified
a natural mechanism for the generation of extremely small scalar
masses without fine-tuning. The mechanism exploits the fact that the
spontaneous breaking of the $\UA$ symmetry would lead to an exactly
massless Goldstone boson in absence of the gauge interactions. The
axial anomaly in the gauge sector then endows a small mass to this
boson.  Owing to the highly different RG behavior of the scalar and
the gauge sector, the scale of P\xsb differs generically from the
scale of nonperturbative gauge effects by many orders of magnitude.
This leads to an exponential suppression of the influence of the axial
anomaly and thus to an exponentially small but nonzero scalar mass.

For theories with a fundamental scalar, the question arises as to
whether our technique of fermion-boson translation is capable of
describing all possible mesonic degrees of freedom.  Let us first look
at two extreme situations. For a large negative renormalized scalar
mass term, perturbation theory applies: there is a fundamental scalar,
and separately propagating meson states may not exist - similar to a
very heavy top quark. For a positive renormalized mass term, the
fundamental scalar decouples from the low-energy sector in
perturbation theory. The low-energy sector then is QCD without
scalars, as in our picture.  The transition is less obvious: in the
region where the fundamental scalar mass would perturbatively be of
the order of the strong interaction scale, there is a strong mixing
between operators corresponding to fundamental and composite scalars.

In principle, in a situation with mixing the propagator in the scalar
sector may have one or several pole-like structures that can be
associated with particle excitations. Our truncation cannot fully
resolve this issue, since, by construction, it follows the flow of
only one pole in the propagator.  Our investigation shows the
consistency of a picture with only one pole.  If the true physical
situation had two poles our truncation would follow the flow of the
lowest mass. We see, however, no indication that a second pole
actually exists. {Nevertheless, it seems worthwhile to discuss the
possible implications of a second scalar ``pole'' for the issue of
universality classes. First of all, for $\bar{m}^2_{\Lambda}$ larger
than but not in the immediate vicinity of $\bar{m}^2_{\text{c}}$, a
``second pole'' could correspond only to an additional heavy scalar
particle. This would decay at a high rate into the QCD mesons, since
no quantum numbers forbid such a decay. (One expects at most a
resonance rather than a true pole.) Furthermore, effects from the
exchange of such a heavy scalar resonance would be suppressed by
inverse powers of the mass and therefore play no role for the
low-energy theory. This is what one usually understands by ``QCD
universality class'', (a notion that is not thought to resolve the
detailed short-distance physics).

This issue becomes more interesting when $\bar{m}^2_{\Lambda}$ is
fine-tuned to the immediated vicinity of $\bar{m}^2_{\text{c}}$. In
this case, we approach the {\em boundary} of the QCD universality
class. We emphasize that this boundary is not uniquely defined in
terms of the symmetries and particles characteristic for the QCD
universality class. Considering the QCD universality class from the
viewpoint of a larger space of models or parameters, the spectrum of
excitations that are relevant at the boundary can depend on the
direction in parameter space from which the QCD universality class is
approached. Different directions may yields a differend ``number of
poles'' in the boundary region. For this reason, a future more
detailed investigation of this issue would be quite interesting.}

We stress that all of our main conclusions can be drawn
from a mere perturbative knowledge of the gauge sector which is well
under control. In a broader sense, the pure QCD sector in our work can
be regarded as a particular example for possible other
(nonperturbatively) renormalizable theories leading to fermionic
self-interactions in scalar channels.

Let us finally discuss our findings from a different perspective,
concentrating on the scalar sector. Scalar fields are known to lead to
profound problems in quantum field theory for two reasons: triviality
and (un-)naturalness. Triviality tells us that an interacting scalar
theory requires a UV cutoff which cannot be removed without switching
off the interaction. Therefore, whenever we see a scalar quantum field
at some low scale, we know that there must be new physics at a higher
scale. The problem of naturalness tells us that it is difficult to
achieve a large separation of scales for models with interacting
scalar fields without fine-tuning.

Our formulation has the potential to solve both problems. A first
example can be given within the QCD universality class. Although from
a QCD perspective, the scalar field could be regarded as purely
auxiliary, nothing prevents us from considering it as fundamental,
since the concepts of compositeness and fundamentality are
interchangeable from the viewpoint of our flow equation with field
transformations. We showed in detail that ``standard'' QCD at low
energies is indistinguishable from QCD with a fundamental scalar, as
long as the latter system is in the QCD universality class. In this
way, we can circumvent triviality by starting in the UV from a scalar
field theory without self-interaction and Yukawa coupling for which
the continuum limit can be taken trivially. The scalar interactions
are induced by quantum fluctuations. In this construction, the system
is always in the QCD universality class, and therefore inherits the
number of relevant and marginal operators form QCD.  In particular, the
scalar mass term is not a relevant operator, so that no naturalness or
fine-tuning problems arise in and from the scalar sector.
Alternatively, we could also follow the bound-state fixed point to
$k\to\infty$, where it presumably becomes an exact fixed point even
beyond our truncation. (The $\beta$ function for the running gauge
coupling vanishes, owing to asymptotic freedom in this limit.)

Perhaps more interesting is a second possibility in the P\xsb
class. Let us consider for $k\to \infty$ a scalar model with
$Z_\phi\to 0$, $\lambda_\phi\to 0$ and $\bar{m}^2$, $\bar{h}$ chosen
such that $\te$ corresponds to the fixed point $\te_1^\ast$. This model
has an alternative interpretation as a model with four-fermion
interactions (and no scalar field). Both the gauge coupling and the
critical four fermion coupling
\begin{equation}
\bar{\lambda}_\sigma^\ast=\frac{1}{2\te_1^\ast k^2} \label{Con1}
\end{equation}
vanish for $k\to \infty$. If this fixed point persists beyond our
truncation, it defines a nonperturbatively renormalizable theory
\cite{Kondo:1992sq}. For lower $k$ a nonzero $\lambda_\phi$ is
generated by the flow and we end up with a theory that effectively
looks like a model with an interacting fundamental scalar field. This
scalar field can give mass to the quarks by P\xsb independently of the
strong interactions, in analogy to the Higgs scalar. The triviality
problem could be solved in this case -- but not the naturalness
problem, since we expect a relevant parameter corresponding to the
scalar mass term.

This discussion sheds new light on the continuous transition between
the P\xsb and QCD universality classes. In the language of statistical
physics, it can be considered as a type of crossover between the
``fundamental fixed point'' $\te_1^\ast$ and the ``bound-state fixed
point'' $\te_2^\ast$. As a particularity, the gauge coupling is a
marginal parameter for both fixed points. The scale where it becomes
strong sets the lowest possible scale for the effective fermion
masses. 

Quite generally, the existence of a bound-state-like fixed point leads
to a mechanism with a naturally small scalar mass. In a sense, this is
a realization of earlier ideas of a large anomalous mass dimension for
the scalar field or ``self-organized criticality''
\cite{Wetterich:1988qu}. It would be interesting to
know if a similar mechanism could contribute to an
understanding of electroweak symmetry breaking which occurs at a
characteristic scale hundreds of times bigger as compared to QCD.

\section*{Acknowledgement}

The authors thank J.~Jaeckel for valuable discussions. H.G.
acknowledges financial support by the Deutsche Forschungsgemeinschaft
under contract Gi 328/1-1.

\section*{Appendix}

\renewcommand{\thesection}{\mbox{\Alph{section}}}
\renewcommand{\theequation}{\mbox{\Alph{section}.\arabic{equation}}}
\setcounter{section}{0}
\setcounter{equation}{0}

\section{Threshold functions}
\label{threshold}

The regularization scheme dependence induced by the cutoff function
$R_k$ is carried by the threshold functions $l$ and $m$. Let us
represent the cutoff functions in the scalar, fermion and gauge sector
by
\begin{equation}
 R_k^\phi(q^2)=Z_{\phi}\, q^2 r(y), \quad 
R_k^\psi(q)=-{\Zy}\fss{q}\, r_{\text{F}}(y), \quad 
\bigl(R_k^A(q)\bigr)_{\mu\nu}\!={\ZF} q^2 r(y)
\left(\!g_{\mu\nu} -\left(\!1-\frac{1}{\xi}\right) \frac{q_\mu q_\nu}{q^2} 
  \!\right)\!, \label{A.0}
\end{equation}
where $y=q^2/k^2$, and $r$ and $r_{\text{F}}$ denote dimensionless
cutoff shape functions. Furthermore, it is useful to introduce the
inverse average propagators $P(x)=x(1+r(x/k^2))$ and
$\PF(x)=x(1+\rF(x/k^2))^2$, where $x=q^2$. 

Most of the threshold functions given above are defined in
Appendix A of \cite{Berges:2000ew}. The ones which cannot be found
therein are marked with a tilde. These can be defined as follows:
\begin{eqnarray}
\text{}\!\!\!\!\!\!\!
&&\tilde{m}_{1,1}^{(\text{FB}),d}(w_{\text{F}},w_{\text{B}}
   ;\eta_\psi;\eta_\phi)\nonumber\\
\text{}\!\!\!\!\!\!\!
&&\quad=
-\frac{1}{2} k^{4-d} \int_0^\infty dx\, x^{d/2-1} \,\patt \left[
  \frac{1+\rF(x/k^2)}{\PF(x)+k^2\wF}
  \frac{1}{P(x)+k^2\wB} \right], \label{T1}\\
\text{}\!\!\!\!\!\!\!
&&\tilde{l}_{1,2}^{(\text{FB}), d}(\wF,\wB;\eta_\psi,\eta_{\text{B}})
  \nonumber\\
\text{}\!\!\!\!\!\!\!
&&\quad= -\frac{1}{2} k^{6-d}\int_0^\infty dx\, x^{d/2-1} \,\patt \left[
 \frac{\PF(x)}{(\PF(x)+k^2\wF)^2 }
    \frac{1}{(P(x)+k^2 \wB)^2}\right], \label{T2}\\
\text{}\!\!\!\!\!\!\!
&&\tilde{l}_{1,1,1}^{(\text{FBB}),
  d}(\wF,\wB{}_1,\wB{}_2;\eta_\psi,\eta_{\text{B}}) \nonumber\\
\text{}\!\!\!\!\!\!\!
&&\quad= -\frac{1}{2} k^{6-d}\int_0^\infty dx\, x^{d/2-1} \,\patt \left[
 \frac{\PF(x)}{(\PF(x)+k^2\wF)^2 }
    \frac{1}{P(x)+k^2 \wB{}_1}
    \frac{1}{P(x)+k^2 \wB{}_2}   \right], \label{T3}
\end{eqnarray}
where $\eta_{\text{B}}$ denotes one of the anomalous dimensions of the
bosonic propagators under consideration, $\eta_\phi$ or $\etaF$ in our
case. The derivative $\patt$ acts on the $k$ dependence of the cutoff
function only (for an explicit representation of $\patt$, see
\cite{Berges:2000ew}). Some relations among the threshold functions
are given by
\begin{eqnarray}
\tilde{l}_{1,1,1}^{(\text{FBB}),
  d}(\wF,\wB,\wB;\eta_\psi,\eta_{\text{B}})&\equiv&
\tilde{l}_{1,2}^{(\text{FB}),
  d}(\wF,\wB;\eta_\psi,\eta_{\text{B}}),\nonumber\\
\tilde{l}_{1,2}^{(\text{FB}), d}(\wF=0,\wB;\eta_\psi,\eta_{\text{B}}) &=&
{l}_{1,2}^{(\text{FB}),
  d}(\wF=0,\wB;\eta_\psi,\eta_{\text{B}}). \label{T4}
\end{eqnarray}
For our numerical computations, we use the linear cutoff functions
proposed in \cite{Litim:2001up} ($y=q^2/k^2$),
\begin{equation}
r(y)=\left(\frac{1}{y}-1 \right)\theta(1-y), \quad
r_{\text{F}}(y)=\left(\frac{1}{\sqrt{y}} -1\right) \theta(1-y),
 \label{o1}
\end{equation}
for which all integrals listed above can be performed analytically,
yielding in the present context:
\begin{eqnarray}
\tilde{m}_{1,1}^{(\text{FB}),d}(\wF,0;\eta_\psi,\eta_F)&=&
\frac{2}{d-1}\, \frac{1}{1+\wF} \left[ \frac{1}{2} \left(1+
    \frac{d}{2} \eta_\psi\right)-\frac{\eta_F}{d+1} +
  \frac{(1-\frac{d}{2} \eta_\psi )}{1+\wF} \right],\label{T5} \\
\tilde{l}_{1,2}^{(\text{FB}),d}(\wF,0;\eta_\psi,\eta_F)&=&
\frac{2}{d} \frac{1}{(1+\wF)^2} \left[ \left( \!1-\frac{2\eta_F}{d+2} +
    \frac{\eta_\psi}{d+1}\! \right) + \frac{2}{1+\wF} \left(\!
    1-\frac{\eta_\psi}{d+1}\! \right) \right]\!, \nonumber \\
\tilde{l}_{1,1,1}^{(\text{FBB}),
  d}(\wF,w_1,w_2;\eta_\psi,\eta_\phi)\!\!&=&\!\!
\frac{2}{d} \frac{1}{(1+\wF)^2(1+w_1)(1+w_2)} \nonumber\\
&&\!\!\times\!\left[\! \left(\!
    \frac{1}{1\!+\!w_1} + \frac{1}{1\!+\!w_2}\!\right)\!
  \left(\!1-\frac{\eta_\phi}{d\!+\!2}\! \right)\!
+ \! \left(\!\frac{2}{1\!+\!\wF}-1\!\right)\!
\left(\!1-\frac{\eta_\psi}{d\!+\!1} 
  \!\right)\!\right]\!. \nonumber
\end{eqnarray}
The representations of all other threshold functions for the linear
cutoff can be looked up in \cite{Hoefling:2002hj}.

\section{Nonperturbative running of the gauge coupling}
\label{beta}

The infrared quantities serving as ``physical observables'' in the
present work, such as the constituent quark mass or the eta boson
mass, depend on the {way we model the effective gauge coupling in
the nonperturbative domain in our truncation}. In order to gain more
insight into this dependence, we study different gauge coupling
$\beta$ functions proposed in the literature in this appendix. Here we
focus on theories within the QCD universality class which are
sensitive to the infrared physics of the gauge sector.

In Sect.~\ref{numerics}, we used a $\beta$ function with accurate
two-loop behavior and an IR fixed-point at $\alpha_\ast=2.5$. We
denote this $\beta$ function defined in Eq.\re{3.2} serving as a
reference as $\beta_{\text{Ref}}$ in the following.  Such $\beta$
functions with a fixed point of the gauge coupling in the infrared
have a long tradition in the literature and have frequently been
discussed from a phenomenological viewpoint \cite{Eichten:1974af}.
Furthermore, some theoretical evidence for the existence of such a
fixed point has been collected in certain nonperturbative
approximation schemes. However, due to the lack of a unique
nonperturbative definition of the gauge coupling and due to an
inherent regularization scheme dependence of the $\beta$ function, a
comparison of different theoretical approaches and a connection to
phenomenology is difficult to make. Here we take a pragmatic point of
view and use the various running couplings as effective ones which are
implicitly defined by their use in our approach.

Recently, an actual nonperturbative computation of the running
coupling has been set up in the framework of truncated Schwinger-Dyson
equations in Landau gauge \cite{vonSmekal:1997is}, revealing an
infrared fixed point; these results also receive some support from
lattice calculations \cite{Bonnet:2001uh}. For our purposes, we use
the representation given in \cite{Fischer:2002hn} for the running
coupling,
\begin{eqnarray}
g_{\text{SDE}}^2(x)&=&\frac{4\pi\,
  \alpha_{\ast,\text{SDE}}}{\ln(e+a_1x^{a_2}+c_1 
  x^{c_2})}, \quad \text{where}\,\, \alpha_{\ast,\text{SDE}}=2.972,
  \label{B.1} 
\end{eqnarray}
and $a_1=5.292$GeV${}^{-2a_2}$, $a_2=2.324$,
$c_1=0.034$GeV${}^{-2c_2}$, $c_2=3.169$. This coupling is also
normalized to the standard value at the $Z$ mass, and we identify
$x=k^2/(\text{GeV})^2$. The $\beta$ function is given by
$\beta_{\text{SDE}}=\pat g_{\text{SDE}}^2$. 

As a second example, we use the running coupling arising from a scheme
called ``Analytic Perturbation Theory'' \cite{Shirkov:1997wi} that has
been devised for enforcing analyticity properties of the coupling in the
time-like and space-like (Euclidean) region. For our numerical
routine, we use the approximate (but two-loop accurate) representation
\begin{eqnarray}
g_{\text{APT}}^2(x)&=&\frac{(4\pi)^2}{b_0} \left( \frac{1}{l_2(x)} +
  \frac{1}{1-\exp[l_2(x)]} \right), \\
&&l_2(x) =\ln x +\frac{b_1}{b_0^2} \ln \sqrt{ \ln^2 x +4\pi^2},
  \label{B.2}
\end{eqnarray}
where we identify $x=k^2/(1349\text{MeV})^2$, so that this coupling is
also normalized at the $Z$ mass. In the infrared $k\to0$, the coupling
tends to the fixed point
$\alpha_{\ast,\text{APT}}=\frac{4\pi}{b_0}\simeq 1.22$ for $\Nc=3$ and
$N_{\text{f}}=1$.

As a third example, we use a calculation of the running coupling based
on a truncated flow equation that also revealed an infrared fixed
point $\alpha_{\ast,\text{FE}}$ \cite{Gies:2002af}. The corresponding
$\beta_{\text{FE}}$ function was obtained as an extensive multiple
integral which we will not display here. Since this result holds for
pure gauge theory, we incorporate one quark flavor in a ``quenched''
approximation by adding the fermionic part of the two-loop $\beta$
function to the pure gauge result. This leads to an infrared
fixed-point value of $\alpha_{\ast,\text{FE}}\simeq3.43\pm0.01$, where
the theoretical error arises from an incompletely resolved color
structure in \cite{Gies:2002af}. We would like to point out that the
definition of the running coupling used in \cite{Gies:2002af} agrees
with the one of the present work.

As a simple example for a running coupling which does not tend to an
infrared fixed point, we employ a class of $\beta$ functions that
correspond to anomalous dimensions of the gauge field $\eta_F$ which
become constant for $k\to 0$. This is realized by the choice
\begin{equation}
\beta_{\eta_\ast}=-2\left(b_0\, \frac{g^4}{16\pi^2}+ b_1\,
\frac{g^6}{(16\pi^2)^2} \right) \left[
1-\exp\left(-\frac{(16\pi^2)^2}{2 b_1 g^4} (-\eta_\ast)\right)\right],
\label{B.3}
\end{equation}
so that in fact $\eta_F=\beta_{\eta_\ast}/g^2\to\eta_\ast$ for $k\to
0$. For negative $\eta_\ast$, the running coupling increases $\sim
(1/k)^{|\eta_\ast|}$ for $k\to 0$. As explicit examples, we choose
$\eta_\ast=-0.1$ and $\eta_\ast=-0.5$ for the numerical analysis.

The results of the numerical integration of the flow equations are
collected in Tab.~\ref{tab}. For the various $\beta$ functions denoted
in the first column, we listed the transition scale $k_{\text{\xsb}}$
into the \xsb regime and the generated fermion mass $m_{\text{f}}$ in
the next two columns. These results refer to calculations without
axial anomaly and instanton-mediated interactions, similarly to
Sect.~\ref{numerics}. In the last two columns the fermion mass with
instanton contribution and the mass of the eta boson are given.

Obviously, the quantities $k_{\text{\xsb}}$ and $m_{\text{f}}$ in the
calculation without axial anomaly are roughly correlated. Furthermore,
$\beta_{\text{APT}}$ and $\beta_{\eta_\ast=-0.1}$ lead to small values
for $k_{\text{\xsb}}$ and $m_{\text{f}}$, since both approach larger
values of the coupling only very slowly. The fermion and eta boson
masses including the axial anomaly are not strictly correlated with
the former quantities. The running of the gauge coupling enters these
quantities over a wider range of scales, since sizable instanton
contributions can already arise while the bound-state fixed point is
still present. 

Nevertheless, our main observation is that the overall qualitative
picture of the approach to \xsb$\!\!$, even in the nonperturbative
domain, is rather independent of the details of the gauge sector
{in our truncation}. On
the one hand, a strong gauge coupling $g^2>g_{\text{D}}^2$ is all that
is needed to trigger \xsb$\!\!$; on the other hand, fermion decoupling
cuts off any strong influence of the running coupling in the deep
infrared.  Quantitative results, of course, depend strongly on the
flow of the coupling between the scale at which $g^2=g_{\text{D}}^2$
and the scale of fermion decoupling. This mainly concerns the overall
scale, whereas mass ratios like $m_\eta/m_{\text{f}}$ turn out to be
more robust.

\begin{table}
\begin{center}
\label{tab}
\begin{tabular}[t]{|l||r|r|r||r|r|}
\hline
$\beta$ function & $m_{\text{f}}/$MeV & $m_{\eta}/$MeV
&$m_{\eta}/m_{\text{f}}$ & $k_{\text{\xsb}}/$MeV 
& $\tilde{m}_{\text{f}}/$MeV \\
\hline
\hline
$\beta_{\text{Ref}}$       & 1765 & 4438 & 2.5 & 423 & 371\\
\hline
$\beta_{\text{SDE}}$       & 1777 & 4226 & 2.4 & 457 & 427\\
\hline
$\beta_{\text{APT}}$       &  563 & 1990 & 3.5 &  4 &   3 \\
\hline
$\beta_{\text{FE}} $       &  903$\pm$2 &
2117$\pm$1 & 2.3 & 243$\pm$2 & 241$\pm$3 \\
\hline
$\beta_{\eta_{\ast}=-0.1}$ &  513 & 1793 & 3.5 &  11 &   8 \\
\hline
$\beta_{\eta_{\ast}=-0.5}$ & 1946 & 4534 & 2.3 & 394 & 395 \\
\hline
\end{tabular}
\end{center}
\caption{Characteristic masses $m_{\text{f}}$ and $m_\eta$ for various
  nonperturbative $\beta$ functions for the strong gauge coupling. The
  main uncertainty concerns the overall scale, whereas the ratio
  $m_\eta/m_{\text{f}}$ is relatively robust. We also show the scale
  of transition to \xsb and the fermion mass $\tilde{m}_{\text{f}}$ in
  absence of instanton effects.}
\label{thetable}
\end{table}

\section{Flow equations with axial anomaly in the broken regime}
\label{nuflow}
Here we collect the flow equations for the various couplings in the
broken regime, including the contributions arising from the axial
anomaly. Let us begin with the scalar and fermion anomalous
dimensions:
\begin{eqnarray} 
\eta_\phi&=&4 v_4\, \kappa\lambda_\phi^2\,
   m_{2,2}^4(\case{\nu}{2\sqrt{\kappa}},\case{\nu}{2\sqrt{\kappa}}
   +2\kappa\lambda_\phi;\eta_\phi) \nonumber\\ 
&&+4\Nc v_4\, h^2\left[ m_4^{(\text{F}),4}(\kappa h^2; \eta_\psi)+
   \kappa h^2\, m_2^{(\text{F}),4}(\kappa h^2; \eta_\psi)\right],
   \label{inst1}\\
\eta_\psi&=&2\Cas v_4\, g^2\Bigl[ (3-\xi)\,
   m_{1,2}^{(\text{FB}),4}(\kappa h^2,0;\eta_\psi,\etaF)- 3(1-\xi)\,
   \tilde{m}_{1,1}^{(\text{FB}),4}(\kappa h^2,0;\eta_\psi,\etaF) \Bigr]
\nonumber\\
&&+v_4\, h^2 \bigl[m_{1,2}^{(\text{FB}),4}(\kappa
   h^2,\case{\nu}{2\sqrt{\kappa}}+2\kappa
      \lambda_\phi;\eta_\psi,\eta_\phi)  
   +m_{1,2}^{(\text{FB}),4}(\kappa h^2,\case{\nu}{2\sqrt{\kappa}}
      ;\eta_\psi,\eta_\phi) \bigr]. \label{inst2}
\end{eqnarray}
Including the appropriately adjusted fermion-boson translation as
outlined in Sect.~\ref{anomaly}, the flow equations for the minimum of
the scalar potential and the scalar self-interaction read:
\begin{eqnarray}
\pat\kappa&=&-(2+\eta_\phi)\kappa \nonumber\\
&&+2 v_4
   \frac{\lambda_\phi}{\lambda_\phi+\frac{\nu}{4\kappa^{3/2}}} 
   \bigl[ l_1^4 (\case{\nu}{2\sqrt{\kappa}};\eta_\phi) 
      + 3 l_1^4(\case{\nu}{2\sqrt{\kappa}}+2\kappa\lambda_\phi;
                  \eta_\phi)\bigr] -8\Nc v_4\, h^4\,
   l_1^{(\text{F}),4}(\kappa h^2;\eta_\psi) \nonumber\\
&& +\frac{2({\kappa\lambda_\phi-\frac{\nu}{2\sqrt{\kappa}}})} 
   {(\lambda_\phi+\frac{\nu}{4\kappa^{3/2}})h^2}
  \left(\!1-\kappa\lambda_\phi+\frac{\nu}{2\sqrt{\kappa}}\right)  
  \label{inst3}\\
&&\qquad\times  \left(1+\left(1-\kappa\lambda_\phi+\frac{\nu}{2\sqrt{\kappa}}
        \right) Q_\sigma\right) \bigl( \beta_{\lk}^{g^4}\, g^4 
  + \beta_{\lk}^{h^4}\, h^4\bigr), \nonumber\\
\pat\lambda_\phi&=& 2\eta_\phi\, \lambda_\phi
   +2v_4\, \lambda_\phi^2\bigl[ l_2^4(\case{\nu}{2\sqrt{\kappa}};\eta_\phi) 
      +9 l_2^4(\case{\nu}{2\sqrt{\kappa}}+2\kappa\lambda_\phi;\eta_\phi)
    \bigr] 
   -8\Nc v_4\, h^4\, l_2^{(\text{F}),4}(\kappa h^2;\eta_\psi)
  \nonumber\\
&&+\frac{4\lambda_\phi}{h^2}
\left[1-2\kappa\lambda_\phi+\frac{\nu}{\sqrt{\kappa}}
  +\left(1-\kappa\lambda_\phi+\frac{\nu}{2\sqrt{\kappa}}\right)^2
  Q_\sigma \right]\bigl( \beta_{\lk}^{g^4}\, g^4 
                          + \beta_{\lk}^{h^4}\, h^4\bigr) \label{inst4}\\
&&+\frac{16\pi^2\lambda_\phi}{\nu h}
   \left(\!\frac{\nu}{2\sqrt{\kappa}}\!-\!\kappa\lambda_\phi\!\right)
    d_0^{\Nc}(f_{\text{c}}/k)\, C_{\text{E}}(\Nc) \left(\!
        \frac{\alpha(k/f_{\text{c}})}
        {\alpha(\bar{\mu})}\!\right)^{-4/b_0}\! \frac{1}{f_{\text{c}}} 
   \left(\!1\!+\! \frac{(-f_{\text{c}}')}{f_{\text{c}}}\, \pat
   (\kappa h^2) \!\right)\!
. \nonumber
\end{eqnarray}
The Yukawa coupling flows in the broken regime according to
 \begin{eqnarray}
\pat h^2&=&(2\eta_\psi+\eta_\phi)\, h^2 -4v_4\, h^4 
   \bigl[ l_{1,1}^{(\text{FB}),4}(\kappa h^2,\case{\nu}{2\sqrt{\kappa}}
      ;\eta_\psi,\eta_\phi) 
   - l_{1,1}^{(\text{FB}),4}(\kappa h^2,\case{\nu}{2\sqrt{\kappa}}
      +2\kappa\lambda_\phi;\eta_\psi,\eta_\phi)\bigr] \nonumber\\
&&-8(3+\xi)\Cas v_4\, g^2 h^2\, l_{1,1}^{(\text{FB}),4}(\kappa h^2,
      0;\eta_\psi,\etaF), \label{inst5}\\
&&+2\left(1-2\kappa\lambda_\phi-\frac{\nu}{\sqrt{\kappa}}
   +\left(1-\kappa\lambda_\phi+\frac{\nu}{2\sqrt{\kappa}}\right)^2
   Q_\sigma \right)\bigl( \beta_{\lk}^{g^4}\, g^4 
  + \beta_{\lk}^{h^4}\, h^4\bigr), \nonumber
\end{eqnarray}
and the flow of the axial anomaly is given by
\begin{eqnarray}
\pat\nu\!\!&=&\!\!-\left(\!3-\frac{\nu_\phi}{2}\!\right) +4\pi^2
  \frac{\kappa\lambda_\phi\!-\!\frac{\nu}{2\sqrt{\kappa}}}{h}
  d_0^{\Nc}(f_{\text{c}}/k)\, 
  C_{\text{E}}\!(\Nc)\left(\!  \frac{\alpha(k/f_{\text{c}})}
    {\alpha(\bar{\mu})}\!\right)^{-4/b_0}\! \frac{1}{f_{\text{c}}}
   \left(\!1\!+\! \frac{(-f_{\text{c}}')}{f_{\text{c}}}\, \pat
   (\kappa h^2)\!\! \right) \nonumber\\
&&+\frac{\nu}{h^2}
  \left[1+\left(1-\kappa\lambda_\phi+\frac{\nu}{2\sqrt{\kappa}}\right)^2
        Q_\sigma\right] 
      \bigl( \beta_{\lk}^{g^4}\, g^4 + \beta_{\lk}^{h^4}\, h^4\bigr).
\label{inst6}
\end{eqnarray}
In these equations, the quantity $\beta_{\lk}^{h^4}$ is also modified,
\begin{eqnarray}
&&\beta_{\lk}^{h^4}:=\frac{2}{\Nc} v_4\, 
  \tilde{l}_{1,1,1}^{(\text{FBB}), 4}(\kappa
  h^2,\case{\nu}{2\sqrt{\kappa}},\case{\nu}{2\sqrt{\kappa}} +
     2\kappa\lambda_\phi;\eta_\psi,\eta_\phi), 
  \label{inst7}
\end{eqnarray}
whereas $\beta_{\lk}^{g^4}$ remains the same. 



\begin{thebibliography}{99}
\setlength{\itemsep}{-0.7mm}
{\small
\bibitem{Wetterich:1993yh}
C.~Wetterich,
Phys.\ Lett.\ B {\bf 301}, 90 (1993);
Nucl.\ Phys.\ B {\bf 352}, 529 (1991);
Z.\ Phys.\ C {\bf 48}, 693 (1990).
%
\bibitem{Gies:2001nw}
H.~Gies and C.~Wetterich,
Phys.\ Rev.\ D {\bf 65}, 065001 (2002)
[arXiv:hep-th/0107221].
%
\bibitem{NJL} Y.~Nambu and G.~Jona-Lasinio, Phys.~Rev.~{\bf 122}, 345
  (1961); {\em ibid.} {\bf 124}, 246 (1961).
%
\bibitem{Berges:2000ew}
J.~Berges, N.~Tetradis and C.~Wetterich,
arXiv:hep-ph/0005122.
%
\bibitem{Abbott:1980hw}
L.~F.~Abbott,
Nucl.\ Phys.\ B {\bf 185}, 189 (1981).

\bibitem{Reuter}
M.~Reuter and C.~Wetterich,
Nucl.\ Phys.\ B {\bf 417}, 181 (1994);\\
%
Phys.\ Rev.\ D {\bf 56}, 7893 (1997)
[arXiv:hep-th/9708051];\\
%
F.~Freire, D.~F.~Litim and J.~M.~Pawlowski,
Phys.\ Lett.\ B {\bf 495}, 256 (2000)
[arXiv:hep-th/0009110].
%
\bibitem{Bonini}
M.~Bonini, M.~D'Attanasio and G.~Marchesini,
Nucl.\ Phys.\ B {\bf 421}, 429 (1994)
[arXiv:hep-th/9312114];\\
%
U.~Ellwanger,
Phys.\ Lett.\ B {\bf 335}, 364 (1994)
[arXiv:hep-th/9402077].
%

%
\bibitem{Pawlowski:2001df}
J.~M.~Pawlowski,
Int.\ J.\ Mod.\ Phys.\ A {\bf 16}, 2105 (2001).
\bibitem{Litim:2002ce}
D.~F.~Litim and J.~M.~Pawlowski,
JHEP {\bf 0209}, 049 (2002)
[arXiv:hep-th/0203005].

\bibitem{Jaeckel:2002rm}
J.~Jaeckel and C.~Wetterich,
arXiv:hep-ph/0207094.
%
\bibitem{Jungnickel:1996fp}
D.~U.~Jungnickel and C.~Wetterich,
Phys.\ Rev.\ D {\bf 53}, 5142 (1996)
[hep-ph/9505267].
%
\bibitem{Chivukula:1992pm}
R.~S.~Chivukula, M.~Golden and E.~H.~Simmons,
Phys.\ Rev.\ Lett.\  {\bf 70}, 1587 (1993).
%
%
\bibitem{Hoefling:2002hj}
F.~Hoefling, C.~Nowak and C.~Wetterich,
arXiv:cond-mat/0203588.
%
\bibitem{Aoki:1996fh}
K.~I.~Aoki, K.~i.~Morikawa, J.~I.~Sumi, H.~Terao and M.~Tomoyose,
Prog.\ Theor.\ Phys.\  {\bf 97}, 479 (1997)
[arXiv:hep-ph/9612459].
%
\bibitem{Aoki:2000dh}
K.~I.~Aoki, K.~Takagi, H.~Terao and M.~Tomoyose,
Prog.\ Theor.\ Phys.\  {\bf 103}, 815 (2000)
[arXiv:hep-th/0002038].
%
\bibitem{Litim:2001up}
D.~F.~Litim,
Phys.\ Lett.\ B {\bf 486}, 92 (2000)
[hep-th/0005245];
Phys.\ Rev.\ D {\bf 64}, 105007 (2001)
[arXiv:hep-th/0103195].
%
%
\bibitem{Ellwanger:1995qf}
U.~Ellwanger, M.~Hirsch and A.~Weber,
Z.\ Phys.\ C {\bf 69}, 687 (1996)
[arXiv:hep-th/9506019];\\
%
Eur.\ Phys.\ J.\ C {\bf 1}, 563 (1998)
[arXiv:hep-ph/9606468].
%
\bibitem{Litim:1998qi}
D.~F.~Litim and J.~M.~Pawlowski,
Phys.\ Lett.\ B {\bf 435}, 181 (1998)
[arXiv:hep-th/9802064].
%
\bibitem{'tHooft:fv}
G.~'t Hooft,
Phys.\ Rev.\ D {\bf 14}, 3432 (1976)
[Erratum-ibid.\ D {\bf 18}, 2199 (1978)];\\
M.~A.~Shifman, A.~I.~Vainshtein and V.~I.~Zakharov,
Nucl.\ Phys.\ B {\bf 163}, 46 (1980);
E.~V.~Shuryak,
Nucl.\ Phys.\ B {\bf 203}, 93 (1982).
%
\bibitem{ilm}
For recent reviews, see
T.~Schafer and E.~V.~Shuryak,
Rev.\ Mod.\ Phys.\  {\bf 70}, 323 (1998)
[arXiv:hep-ph/9610451];\\
D.~Diakonov,
arXiv:hep-ph/0212026.
\bibitem{Pawlowski:1996ch}
J.~M.~Pawlowski,
Phys.\ Rev.\ D {\bf 58}, 045011 (1998)
[arXiv:hep-th/9605037].
%
\bibitem{Kondo:1992sq}
K.~i.~Kondo, M.~Tanabashi and K.~Yamawaki,
Prog.\ Theor.\ Phys.\  {\bf 89}, 1249 (1993)
[arXiv:hep-ph/9212208];\\
%
%
K.~I.~Kubota and H.~Terao,
Prog.\ Theor.\ Phys.\  {\bf 102}, 1163 (1999)
[arXiv:hep-th/9908062];\\
%
M.~Reenders,
Phys.\ Rev.\ D {\bf 62}, 025001 (2000)
[arXiv:hep-th/9908158].
%
\bibitem{Wetterich:1988qu}
C.~Wetterich,
Phys.\ Lett.\ B {\bf 209}, 59 (1988);\\
S.~Bornholdt and C.~Wetterich,
Phys.\ Lett.\ B {\bf 282}, 399 (1992).
%
\bibitem{Eichten:1974af}
E.~Eichten {\em et al.},
Phys.\ Rev.\ Lett.\  {\bf 34}, 369 (1975)
[Erratum-ibid.\  {\bf 36}, 1276 (1975)];\\
%
T.~Barnes, F.~E.~Close and S.~Monaghan,
Nucl.\ Phys.\ B {\bf 198}, 380 (1982);\\
%
S.~Godfrey and N.~Isgur,
Phys.\ Rev.\ D {\bf 32}, 189 (1985);\\
%
A.~C.~Mattingly and P.~M.~Stevenson,
Phys.\ Rev.\ Lett.\  {\bf 69}, 1320 (1992);
[arXiv:hep-ph/9207228];\\
%
A.~C.~Aguilar, A.~Mihara and A.~A.~Natale,
arXiv:hep-ph/0208095.
%
\bibitem{vonSmekal:1997is}
L.~von Smekal, R.~Alkofer and A.~Hauck,
Phys.\ Rev.\ Lett.\  {\bf 79}, 3591 (1997)
[arXiv:hep-ph/9705242];\\
%
Annals Phys.\  {\bf 267}, 1 (1998)
[Erratum-ibid.\  {\bf 269}, 182 (1998)]
[arXiv:hep-ph/9707327];\\
%
D.~Atkinson and J.~C.~Bloch,
Phys.\ Rev.\ D {\bf 58}, 094036 (1998)
[arXiv:hep-ph/9712459];\\
%
D.~Zwanziger,
arXiv:hep-th/0109224;\\
%
%
C.~Lerche and L.~von Smekal,
arXiv:hep-ph/0202194.
%
\bibitem{Bonnet:2001uh}
F.~D.~Bonnet, P.~O.~Bowman, D.~B.~Leinweber, A.~G.~Williams and J.~M.~Zanotti,
Phys.\ Rev.\ D {\bf 64}, 034501 (2001)
[arXiv:hep-lat/0101013];\\
%
J.~R.~Bloch, A.~Cucchieri, K.~Langfeld and T.~Mendes,
arXiv:hep-lat/0209040.
%
\bibitem{Fischer:2002hn}
C.~S.~Fischer and R.~Alkofer,
arXiv:hep-ph/0202202.
%
\bibitem{Shirkov:1997wi}
D.~V.~Shirkov and I.~L.~Solovtsov,
Phys.\ Rev.\ Lett.\  {\bf 79}, 1209 (1997)
[arXiv:hep-ph/9704333];\\
%
Theor.\ Math.\ Phys.\  {\bf 120}, 1220 (1999)
[Teor.\ Mat.\ Fiz.\  {\bf 120}, 482 (1999)]
[arXiv:hep-ph/9909305].
%
\bibitem{Gies:2002af}
H.~Gies,
Phys.\ Rev.\ D {\bf 66}, 025006 (2002)
[arXiv:hep-th/0202207].
}
\end{thebibliography}
\end{document}